\documentclass[fleqn,10pt]{wlscirep}
\usepackage{graphicx}
\usepackage{dcolumn}
\usepackage{bm}
\usepackage{subfigure}
\usepackage{color}

\title{Minimum Copies of Schr\"{o}dinger's Cat State in the Measurement of Multi-Photon Entanglement}

\author[1]{Yiping Lu}
\author[1,*]{Qing Zhao}
\affil[1]{School of Physics, Beijing Institute of Technology, Haidian District, Beijing 100081, P.R. China}

\affil[*]{qzhaoyuping@bit.edu.cn}

\begin{abstract}
Multi-photon entanglement has been successfully studied by many theoretical and experimental groups. However, as the number of entangled photons increases, some problems are encountered, such as, the exponential increase of time necessary to prepare the same number of copies of entanglement states in experiment. In this paper, a new scheme is proposed based on Lagrange multiplier and Feedback, which cuts down the required number of copies of Schr\"{o}dinger's Cat state in multi-photon experiment and keeps the standard deviation error of fidelity unchanged. It reduces five percent of the measuring time of eight-photon Schr\"{o}dinger's Cat state compared with the existing, and also guarantees the low error in fidelity. Additionally, the same approach has also been applied to the simulation of ten-photon entanglement. It reduces the twenty two percent of the copies of ten-photon Schr\"{o}dinger's Cat state compared with the uniform distribution, and the distribution of optimized copies gives better fidelity estimation than the uniform distribution of copies in the case of the present the ten-photon Schr\"{o}dinger's Cat state.
\end{abstract}
\begin{document}

\flushbottom
\maketitle
%
%
\thispagestyle{empty}

\section*{\label{sec:level1}Introduction}
 From fundamental tests of quantum mechanics \cite{EINsTEIN} to quantum teleportation, quantum key distribution, and quantum communications \cite{Lee,Tang,Vallone}, quantum entanglement has wide applications in different areas. Recently, a single photon has been recognized to teleport multiple degrees of freedom simultaneously, which includes spin and orbital angular momentum \cite{Wang}. The Greenberger-Horne-Zeilinger (GHZ) states created in experiment have been
obtained by combining the momentum and polarization \cite{1,2,3,4,5}. Several experiments have been performed to validate multi-photon entanglement \cite{threephotonGHZ,1,2,3,5,39}. An indispensable tool is the entanglement witness for certification of entanglement in these related experiments \cite{43,GO,HR}. Generally, the expectation value of entanglement witness can be evaluated by fidelity \cite{37}. Precise estimation of fidelity requires many identical copies of the prepared state \cite{fidcopy}. On the other hand, the coincidence count rate of multi-photon entangled states decreases exponentially with a linear increase in the number of entanglement photons, which is generated by the phenomenon of a nonlinear process of parametric down-conversion in BBO \cite{PANRMP,8,Crosse,Ghosh,Parigi}. Hence, collecting sufficient copies of multi-photon entanglement state costs much longer time, such as, in eight-photon entanglement it takes $170$ hours to produce the sufficient copies of eight-photon Schr\"{o}dinger's cat (SC) state (See the label of Fig.3 of Ref.\cite{2}). Up to now, a study on ten-photon entanglement or more is inaccessible in experiment since the coincidence count rate of ten-photon entanglement state is lower than 9 counts per hour \cite{2}. It needs nearly three months to prepare sufficient copies, for example, 110, of the ten-photon SC state to certify entanglement according to the current technology. (See Appendix ``Preparation of Ten-photon SC state").

 Discrimination of a quantum state by adaptive process is developed recently. Adaptive process is to split the conventional measurement into several pieces and to choose the current measurement based on the results of previous measurements. The standard of selection is chosen to be minimizing the probability of error \cite{pryde}. The probability is estimated by the known information. Generally the adaptive process is split into two steps. The first step is to get rough information. The second step is to rectify it and get a precise density matrix \cite{D.H.Mahler, thePaperOfGuoLaoshi}.

In this paper, an efficient method is developed to reduce the number of copies of an unknown state to certificate entanglement in the multi-photon experiment. Specifically, the conventional measurement is split into several steps. For each step, the least number and the optimal distribution of identical copies of the unknown state on different measurement settings are calculated by the proposed model. The measurement result from previous steps provides the value of parameters for future steps. In our model, the unknown state is supposed to be a pure SC state or SC state in the presence of noise. Since the entanglement validation of SC state is through fidelity \cite{43}, the optimization introduces fidelity as a criterion. When fidelity is greater than 0.5, the experimentally prepared state is certified to be entangled \cite{43}. The target of optimization is to search for the minimum number of copies of the unknown state that can confirm the error gap of fidelity belonging to a small area; therefore the fidelity interval can be estimated and the minimum number of copies of state is obtained.

\section*{\label{sec:level2} Minimum copies of multi-photon Schr\"{o}dinger's Cat state}

The experimental n-qubit SC state is denoted by a $2^n\times2^n$ density matrix $\rho_{exp}$. Its fidelity with the pure state $|SC\rangle$ is defined as
\begin{eqnarray}
F_{exp}(|SC\rangle)=\langle SC|\rho_{exp}|SC\rangle = {\rm Tr}(\rho_{exp}|SC\rangle\langle SC|).\label{4}
\end{eqnarray}
To calculate the $F_{exp}(|SC\rangle)$, Eq.(\ref{4}) can be written as
\begin{eqnarray}
F_{exp}(|SC\rangle) ={\rm Tr}\left\{\rho_{exp}\left[\frac{1}{2}I-\left(\frac{1}{2}I-|SC\rangle\langle SC|\right)\right]\right\} .\label{FEXP}
\end{eqnarray}
Now setting entanglement witness operator $w$,
\begin{eqnarray}
w=\frac{1}{2}I-|SC\rangle\langle SC|\label{6}
\end{eqnarray}
in Eq.(\ref{FEXP}), we arrive at
\begin{eqnarray}
F_{exp}(|SC\rangle)={\rm Tr}\left[\rho_{exp}\left(\frac{1}{2}I-w\right)\right]=\frac{1}{2}-\langle w\rangle,
\end{eqnarray}
where $\langle w\rangle$ is the expectation of entanglement witness \cite{Guhne2,37}. Hence, $F_{exp}(|SC\rangle)$ can be calculated by evaluating $\langle w\rangle$.
In Eq.(\ref{6}), $|SC\rangle\langle SC|$ is decomposed into the form
\begin{eqnarray}
&&|SC\rangle\langle SC|=\nonumber\\
&&\frac{1}{2}\left[(|H\rangle\langle H|)^{\otimes n}+(|V\rangle\langle V|)^{\otimes n}+\frac{1}{n}\sum_{k=1}^{n} (-1)^{k}M^{\otimes n}_{k\pi/n}\right],\label{DEsc}
\end{eqnarray}
where $M_{k\pi/n}=\cos(k\pi/n)\sigma_{x}+\sin(k\pi/n)\sigma_{y}$ \cite{5,37}. See Appendix ``Entanglement Witness" for more details.

The $n$-qubit SC state requires at least $n+1$ settings to calculate fidelity (see Observation 1 in \cite{37}). Based on Eq.(\ref{DEsc}), the standard deviation of fidelity is deduced,
\begin{eqnarray}
&&\Delta F_{exp}=\sqrt{\frac{1}{4}\frac{P_1(1-P_1)}{t_1}+\frac{1}{n^2}\sum_{j=2}^{n+1}\frac{P_j(1-P_j)}{t_j}}.\label{deltaF1}
\end{eqnarray}
 In Eq.(\ref{deltaF1}), $t_j$ is the total number of copies of n-qubit entanglement state that projected into the $j$th measurement setting, $j=1,2,\cdots, n+1.$ Its value equals the sum of accumulated n-fold coincidence counts in all different bases of the $j$th setting. Here accidental coincidence count is ignored since it is almost zero when $n$ is large. The $P_1$ is equal to the summation of two relative frequencies. One is the case that all qubits are projected into horizontal polarizations $(|H\rangle\langle H|)^{\otimes n}$, the other is the case that all qubits are projected into vertical polarizations $(|V\rangle\langle V|)^{\otimes n}$. Here, the meaning of relative frequency is the ratio between the number of copies of a state projected into a base and the number of copies of the state measured in all the bases belonging to this setting. Similarly, the $P_j$ is the linear combination of relative frequencies of different basis in the $j$th setting. It should be kept in mind that measurement setting means a group of complete basis that copies of a state are projected in the same period of time and relative frequencies being gained simultaneously. The details are shown in the appendix ``Standard Deviation of Fidelity".

We intend to apply fewer copies of the unknown state to estimate fidelity with same accuracy. Let $\epsilon_0$ denote the given upper bound of standard deviation of fidelity since the number of copies of a state required relies on it. Our objective is to use as few copies of the state as possible and, at the same time, to narrow down the fidelity to a small interval. Therefore, the following model is proposed: for $n$-qubit SC state, we have
\begin{eqnarray}
&& {\rm Minimize}\ \sum_{j=1}^{n+1}t_j\  \nonumber\\
&& {\rm subject\ to}\ \nonumber\\
&&\sqrt{\frac{1}{4}\frac{P_1(1-P_1)}{t_1}+\frac{1}{n^2}\sum_{j=2}^{n+1}\frac{P_j(1-P_j)}{t_j}}\leq \epsilon_0.\label{Optimizationproblem}
\end{eqnarray}

 It should be noted that $t_j$ obtained by solving Eq.(\ref{Optimizationproblem}) is sufficiently large, since the larger $t_j$ is, the higher the probability for the result of Eq.(\ref{Optimizationproblem}) to hold, which will be discussed in section of ``Characteristics of optimization of the successful probabilities". Based on numerical results, the solution of Eq.(\ref{Optimizationproblem}) has large $t_j$ in most cases and the probability for the above model is nearly $1$. Following is the analytical solution of Eq.(\ref{Optimizationproblem}) obtained. Let $\epsilon_0^2=\epsilon$, then
 \begin{eqnarray}
&& t_1=\frac{\frac{1}{2}\sqrt{P_1(1-P_1)}\left[\frac{1}{2}\sqrt{P_1(1-P_1)}+\sum_{j=2}^{n+1}\frac{1}{n}\sqrt{P_j(1-P_j)}\right]}{\epsilon}, \nonumber\\
&& t_j=\frac{\frac{1}{n}\sqrt{P_j(1-P_j)}\left[\frac{1}{2}\sqrt{P_1(1-P_1)}+\sum_{j=2}^{n+1}\frac{1}{n}\sqrt{P_j(1-P_j)}\right]}{\epsilon}. \label{optimalt}
\end{eqnarray}

The process to get the Eq.(\ref{optimalt}) can be found in ``Theoretical derivation of minimum copies of multi-photon Schrodinger's Cat state"

\section*{Results}
\subsection*{Direct estimation of fidelity for experimental eight-photon SC state and simulated ten-photon SC state}

The advantage of our method over the existing approaches can be demonstrated by the experiment of eight-photon entanglement. When sufficient copies of eight-photon SC state in the presence of noise ($\rho_{8photons}$) are projected into different settings in experiment, fidelity is calculated by total accumulated coincidence counts on different basis and then the eight-photon entanglement can be verified \cite{2}. In this section, our method is to change the number of copies of prepared SC state ($\rho_{8photons}$) measured in different settings. The results show that total copies of prepared SC state can be saved; while, fidelity precision remains same.

To apply Eq.(\ref{optimalt}) for certificating of eight-photon entanglement in our model, some parameters need to be given, such as $n=8$. Let the prepared eight-photon SC state in experiment be $\rho_{8photons}$, which is a SC state mixed with noise. In order to have optimized results compared with the experiment. The error bound of fidelity, $\epsilon_0$, is set at 0.016, which is exactly the same value as the one used in experiment. This number can be found in the second to last paragraph of Ref.\cite{2}. According to entanglement witness, an eight-photon SC state requires nine settings to determine fidelity uniquely, see equation (2) in Ref.\cite{2}. Let $|H\rangle$ represent horizontal polarization and $|V\rangle$ represent vertical polarization. And define $|+\rangle$$=$$(|H\rangle+ e^{i\theta}|V\rangle)/\sqrt{2}$, $|-\rangle$$=$$(|H\rangle- e^{i\theta}|V\rangle)/\sqrt{2}$. Then the nine measurement settings are defined as

 \begin{eqnarray}
&&  S_{8photons,1}=\{(|H\rangle\langle H|)^{\otimes8}, (|H\rangle\langle H|)^{\otimes7}(|V\rangle\langle V|),\cdots,(|V\rangle\langle V|)^{\otimes8}  \}, \nonumber\\
&&     S_{8photons,2}=\{(|+,\theta\rangle\langle +,\theta|)^{\otimes8}, (|+,\theta\rangle\langle +,\theta|)^{\otimes7}(|-,\theta\rangle\langle -,\theta|),\cdots,\nonumber\\
&& (|-,\theta\rangle\langle -,\theta|)^{\otimes8}, \theta=0  \}, \nonumber\\
&&    S_{8photons,3}=\{(|+,\theta\rangle\langle +,\theta|)^{\otimes8}, (|+,\theta\rangle\langle +,\theta|)^{\otimes7}(|-,\theta\rangle\langle -,\theta|),\cdots,\nonumber\\
&&(|-,\theta\rangle\langle -,\theta|)^{\otimes8}, \theta=\pi/8 \}, \nonumber\\
&&    S_{8photons,4}=\{(|+,\theta\rangle\langle +,\theta|)^{\otimes8}, (|+,\theta\rangle\langle +,\theta|)^{\otimes7}(|-,\theta\rangle\langle -,\theta|),\cdots,\nonumber\\
&& (|-,\theta\rangle\langle -,\theta|)^{\otimes8}, \theta=2\pi/8 \}, \nonumber\\
&&    S_{8photons,5}=\{(|+,\theta\rangle\langle +,\theta|)^{\otimes8}, (|+,\theta\rangle\langle +,\theta|)^{\otimes7}(|-,\theta\rangle\langle -,\theta|),\cdots,\nonumber\\
&&(|-,\theta\rangle\langle -,\theta|)^{\otimes8}, \theta=3\pi/8 \}, \nonumber\\
&&    S_{8photons,6}=\{(|+,\theta\rangle\langle +,\theta|)^{\otimes8}, (|+,\theta\rangle\langle +,\theta|)^{\otimes7}(|-,\theta\rangle\langle -,\theta|),\cdots,\nonumber\\
&& (|-,\theta\rangle\langle -,\theta|)^{\otimes8}, \theta=4\pi/8  \}, \nonumber\\
&&    S_{8photons,7}=\{(|+,\theta\rangle\langle +,\theta|)^{\otimes8}, (|+,\theta\rangle\langle +,\theta|)^{\otimes7}(|-,\theta\rangle\langle -,\theta|),\cdots,\nonumber\\
&&(|-,\theta\rangle\langle -,\theta|)^{\otimes8}, \theta=5\pi/8  \}, \nonumber\\
&&    S_{8photons,8}=\{(|+,\theta\rangle\langle +,\theta|)^{\otimes8}, (|+,\theta\rangle\langle +,\theta|)^{\otimes7}(|-,\theta\rangle\langle -,\theta|),\cdots,\nonumber\\
&&(|-,\theta\rangle\langle -,\theta|)^{\otimes8}, \theta=6\pi/8  \}, \nonumber\\
&&  S_{8photons,9}=\{(|+,\theta\rangle\langle +,\theta|)^{\otimes8}, (|+,\theta\rangle\langle +,\theta|)^{\otimes7}(|-,\theta\rangle\langle -,\theta|),\cdots,\nonumber\\
&&(|-,\theta\rangle\langle -,\theta|)^{\otimes8}, \theta=7\pi/8  \}. \nonumber\\
 \end{eqnarray}

Furthermore, 1305 copies of $\rho_{8photons}$ are prepared in total in experiment \cite{2}. Notice that 1305 is not directly given in the Ref.\cite{2}, but is used to draw the graphes, calculate fidelity in Ref.\cite{2} and provided by the author of that paper. The number can also be roughly calculated by the copies of $\rho_{8photons}$ per hour and the total hours spent. That is $9\times40+9\times25+9\times15\times7=1544$, in which the coincidence counting rate can be found in the 11th paragraph of Ref. \cite{2} and the hours spent for different settings can be found in the label of Figure 3 of Ref. \cite{2}. Accidental coincidence count is very small in eight-photon experiment, therefore it is neglected. A numerical test is performed. Firstly, a set of copies of a quantum state measured at various settings is defined as ``distribution of copies".
 Three different distributions (experimentally applied distribution of copies of $\rho_{8photons}$, optimal distribution of copies of the state, which is obtained from Eq. (\ref{optimalt}), and uniformity distribution of copies of the state) are considered separately, and compared with each other. The number of copies of $\rho_{8photons}$ for each case is listed in Table \ref{tablecopy}. The first column is the tag of setting. The optimal distribution of the copies of $\rho_{8photons}$ calculated by Eq.(\ref{optimalt}) is listed in the last column. Obviously, the total number of copies of $\rho_{8photons}$ required is cut down to 1253, thus 52 copies of $\rho_{8photons}$ (about 5 percent) are saved compared with experiment. Since coincidence count rate is only nearly nine (8.88) per hour (in the 11th paragraph of Ref.\cite{2}), around 5.9 hours could be reduced in the experiment while the same precision of fidelity can be hold.

\begin{table}
\begin{tabular}{|c|c|c|c|}
 \hline
  Setting & Experiment & Uniformity &Optimization \\
  \hline
   $S_{8photons,1}$ & 352 & 145 & 415 \\
   $S_{8photons,2}$  &  200&145 & 106 \\
    $S_{8photons,3}$ &107& 145 & 103  \\
   $S_{8photons,4}$ &100 &145 & 106\\
   $S_{8photons,5}$ &  110 & 145 &103 \\
  $S_{8photons,6}$ & 111 &145 &  108  \\
   $S_{8photons,7}$  & 106&145 & 101 \\
  $S_{8photons,8}$& 116 & 145 &108 \\
   $S_{8photons,9}$ & 103  &145 &  103 \\
  \hline
  Summation & 1305& 1305 &1253\\
  \hline
\end{tabular}
\caption{ Distribution of copies of $\rho_{8photons}$ under different case. The first row represents the number of copies of $\rho_{8photons}$ that is the summation of the number of accumulated coincidence counts projected into all the bases of the first measurement setting. The following eight rows represent the number of copies of $\rho_{8photons}$ that projected into the other eight settings corresponding to $\theta=0, \pi/8,\cdots,7\pi/8$. The last row is the total cost of copies of $\rho_{8photons}$ in the experiment of Ref.\cite{2}, uniform distribution and our optimization. The second column is the cost of number of copies of $\rho_{8photons}$ in the experiment for different setting. The third column represents uniform distribution of copies of $\rho_{8photons}$ in each setting. The last column represents the copies of $\rho_{8photons}$ obtained in the optimization.}\label{tablecopy}
\end{table}

\begin{figure}[!h]
\centering
\includegraphics[width=16cm]{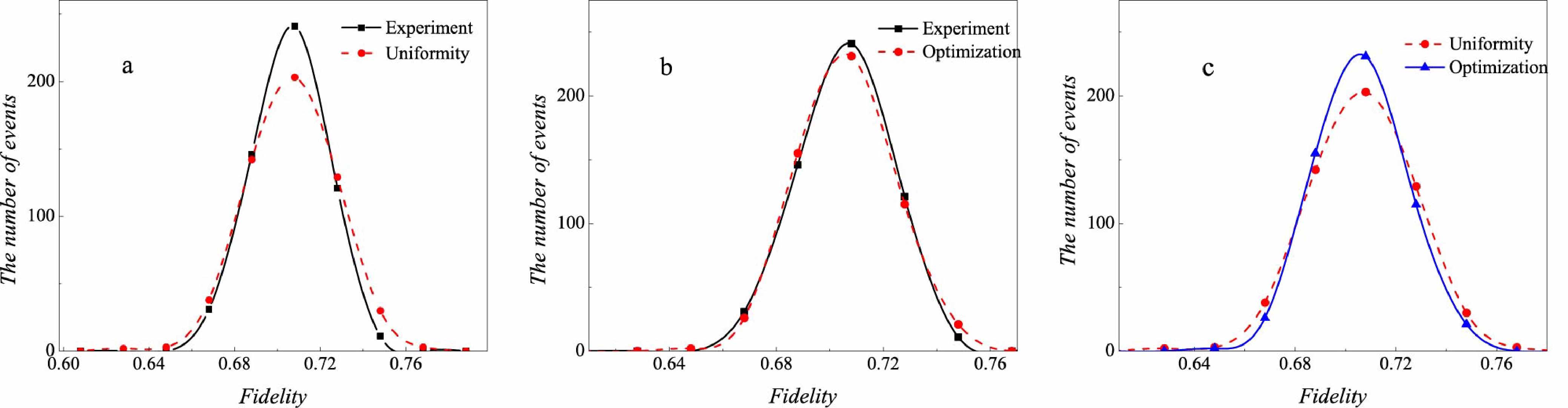}
\caption{a: The number of events versus fidelities for both experimental distribution and uniform distribution. b: The outlines for experiment and optimization, which almost coincide with each other. Optimization only costs 1253 copies of $\rho_{8photons}$, which is a smaller number than the 1305 copies of $\rho_{8photons}$ required by the experiment. c: The number of events versus fidelities for both optimization distribution and uniform distribution.}\label{eightphoto1}
\end{figure}

For each case, fidelity can be calculated from new relative frequencies obtained by simulating the experimental process in computer according to the real precise relative frequencies in different settings. In simulation, the real relative frequency is calculated according to Born's rule. It also requires density matrix to be known in this rule. Fortunately, the density matrix of $\rho_{8photons}$ can be obtained by experimental data and phaselift approach, which will be given in detail in ``optimization of multi-qubit experimental and simulated data via density matrices". Since summation of the real relative frequency of different bases in a same setting is equal to one, the interval between 0 and 1 is divided into $2^8$ sub-intervals and the range for each of the sub-interval is equal to the value of the corresponding relative frequency. And a random number between 0 and 1 is produced with the equal probability for each value between 0 and 1. And the interval it lies in is found and the number of event for this interval is added to one. After producing random numbers with the number of copies of $\rho_{8photons}$ for the setting, different interval gets a different number of event. Then relative frequencies can be calculated. After that simulated fidelity is obtained. We also divide the fidelity range from 0 to 1 into 50 equal portions. Event number is added to one when the calculated fidelity belongs to the corresponding interval. All three situations (experiment, optimization and uniformity) are all repeated for 550 times separately, which means 550 fidelities are calculated. The number of events per interval is accumulated and observed, as shown in Figure \ref{eightphoto1}. Figure \ref{eightphoto1}a shows that: when all 1305 copies of $\rho_{8photons}$ are applied, the experimental results give better estimation of fidelity than the uniform distribution since the height of the outline for uniform distribution on vertical axis direction is lower than the experimental one. The outlines for experiment and optimization are also described, which almost coincide with each other. However, optimization only costs 1253 copies of the $\rho_{8photons}$, which is smaller than the 1305 copies of $\rho_{8photons}$ required by the experiment, as shown in Figure \ref{eightphoto1}b. The Figure \ref{eightphoto1}c demonstrates that optimization is also better than the uniform distribution.

\begin{figure}[!h]
\begin{center}
  \includegraphics[width=8cm]{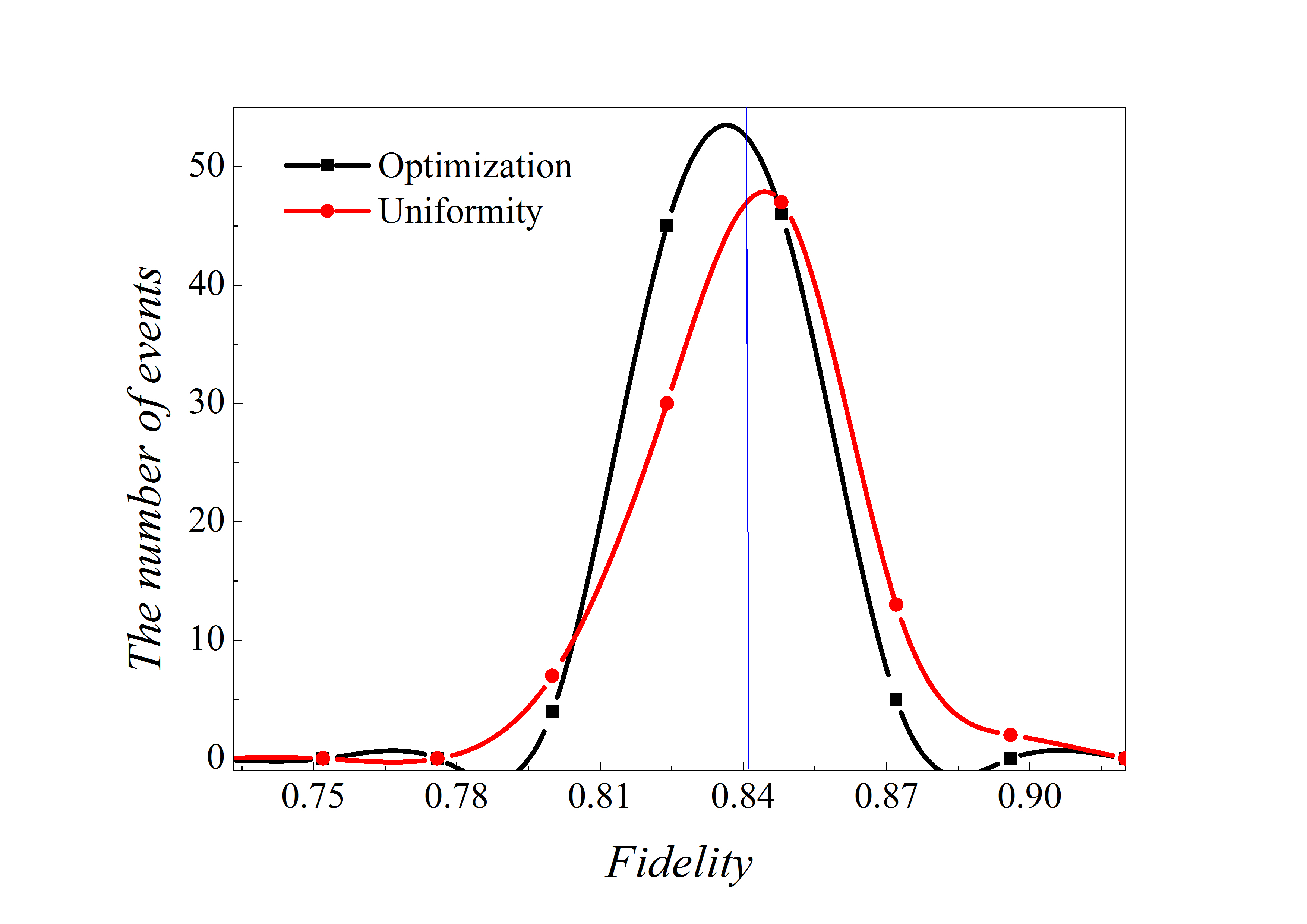}\\
  \caption{ The distribution of events' number of fidelity of ten-photon entanglement state ($\rho_{10photons}$). The blue vertical line represents the fidelity between the simulated state $\rho_{10photons}$ and pure ten-photon SC state, which equals to 0.8414. Different lines are used to connect adjacent points. Black line represents the optimized fidelity distribution and red line represents uniform distribution. It is observed that optimized distribution has more events accumulated near the real fidelity at 0.8414 than uniform distribution.}\label{densitymatriten}
  \end{center}
\end{figure}

At present there is no way to create enough copies of ten-photon SC state to certify entanglement in experiment. Numerical test is produced to estimate fidelity based on a computer created density matrix $\rho_{10photons}$ that its fidelity with pure ten-photon SC state is 0.8414. It is carried out in the situation of uniform distribution in each setting (100 copies of ten-photon SC state for each setting) and optimization. The process is same as eight-photon entanglement. Both cases are all repeated for 100 times separately, then the distributions of fidelities are obtained, as shown in Figure \ref{densitymatriten}. It is observed that 22.45 \% copies of simulated ten-photon SC state $\rho_{10photons}$ can be saved according to Eq.(\ref{optimalt}) compared with uniform distribution on each setting.

Obviously, the optimization yields a better estimation of fidelity with limited copies of state available. The scheme given here is useful in certifying the multi-qubit entanglement state and can be generalized to any state by changing the form of constraint of Eq.(\ref{Optimizationproblem}).

\subsection*{Optimization of multi-qubit experimental and simulated data via density matrices}

 In addition to the direct estimation of fidelity, we also estimate a density matrix first by phaselift \cite{luyiping}, and then calculate fidelity.

The model for calculation of the density matrix is constructed based upon the procedures given in Refs.\cite{James,Banaszek,21, 20, 31, 33, 35, 36}; the noise case is applied, which is in \cite{34} and \cite{luyiping},
 \begin{eqnarray}
&&  {\rm Minimize}\ \sum_{\mu,\nu}|{\rm Tr}(\rho M_{\mu,\nu})-f_{\mu,\nu}|  \nonumber\\
&& {\rm subject\ to}\ \ \rho\geq0, {\rm Tr}(\rho)=1,\label{optidensitymatirx}
\end{eqnarray}
where $\rho$ is density matrix, $M_{\mu,\nu}$ is positive operator valued measure (POVM) in the $\mu$-th bases of the $\nu$-th setting, $f_{\mu,\nu}$ is the relative frequency in the $\mu$-th bases of the $\nu$-th setting.

 The quantum state tomography for three, four and eight-photon entanglement is conducted. When corresponding experimental frequencies $f_{\mu,\nu}$ are put in
 Eq.(\ref{optidensitymatirx}), the density matrix is calculated out. Our objective is to use the least copies of an unknown state to obtain a density matrix close to the real one. The real density matrix $\rho_{exp}$ is approximately obtained with the use of a large number of copies of the state prepared in experiment. Then, $\rho_{exp}$ is applied to gain new frequencies according to Born's rule through the simulation of experiment process on computer. These frequencies are applied to obtain the density matrix $\rho_{re}$. Then many density matrices $\rho_{re}$ are obtained under different number of copies of the state and compared with the $\rho_{exp}$ achieved in the experiment.

\begin{figure}[!h]
\begin{center}
  \includegraphics[width=18cm]{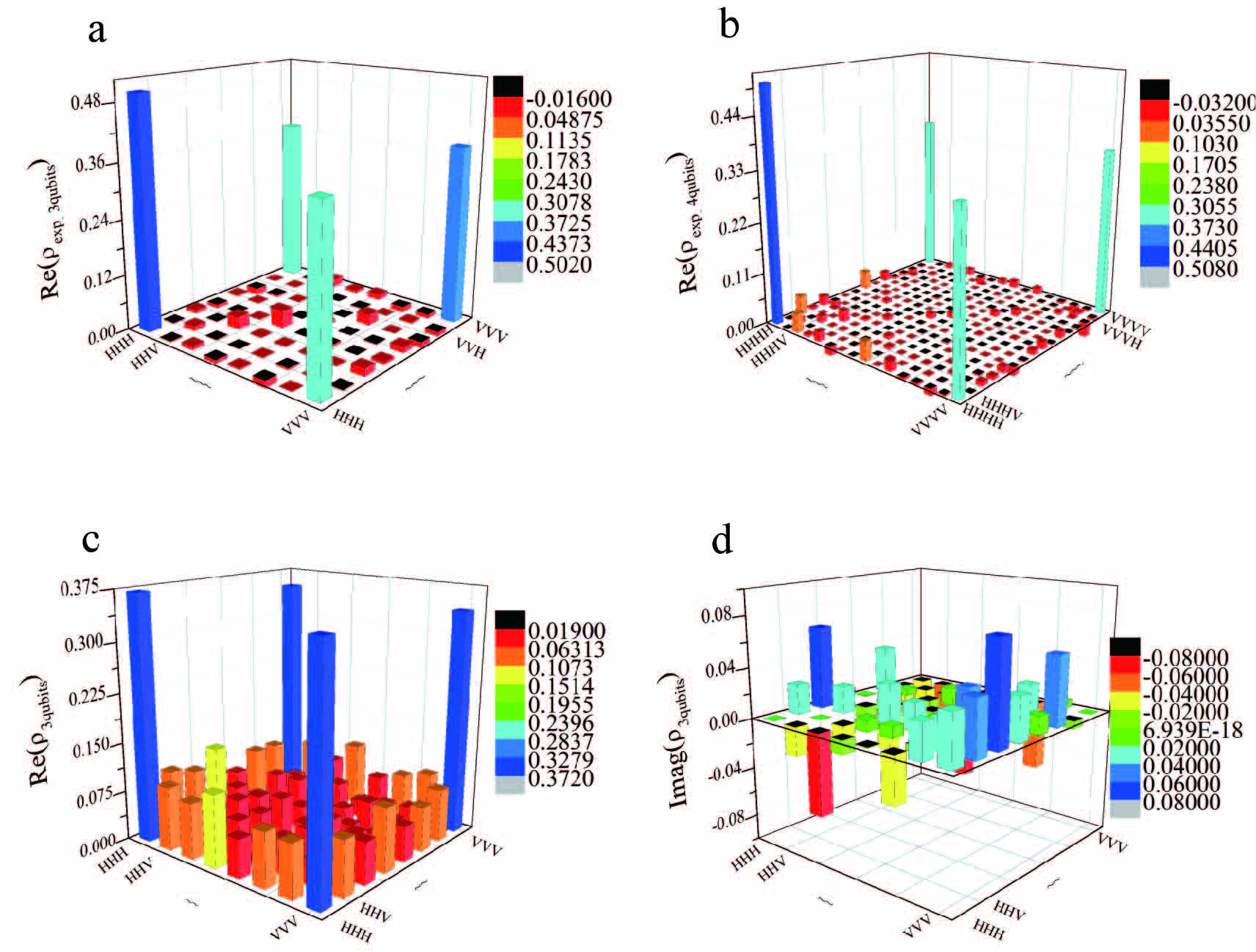}\\
  \caption{ Density matrices ($\rho_{exp_{-}3qubits}$, $\rho_{exp_{-}4qubits}$, $\rho_{3qubits}$). Different colors are applied to represent the value of elements of density matrices. a: The real part of experimental density matrix of three-qubit ($\rho_{exp_{-}3qubits}$). b: The real part of experimental density matrix of four-qubit ($\rho_{exp_{-}4qubits}$). c: Real part of three-photon density matrix ($\rho_{3qubits}$) by random creation. d: Imaginary part of three-photon density matrix ($\rho_{3qubits}$) by random creation. }\label{Figure5789}
  \end{center}
\end{figure}

\begin{figure}[!h]
\begin{center}
  \includegraphics[width=10cm]{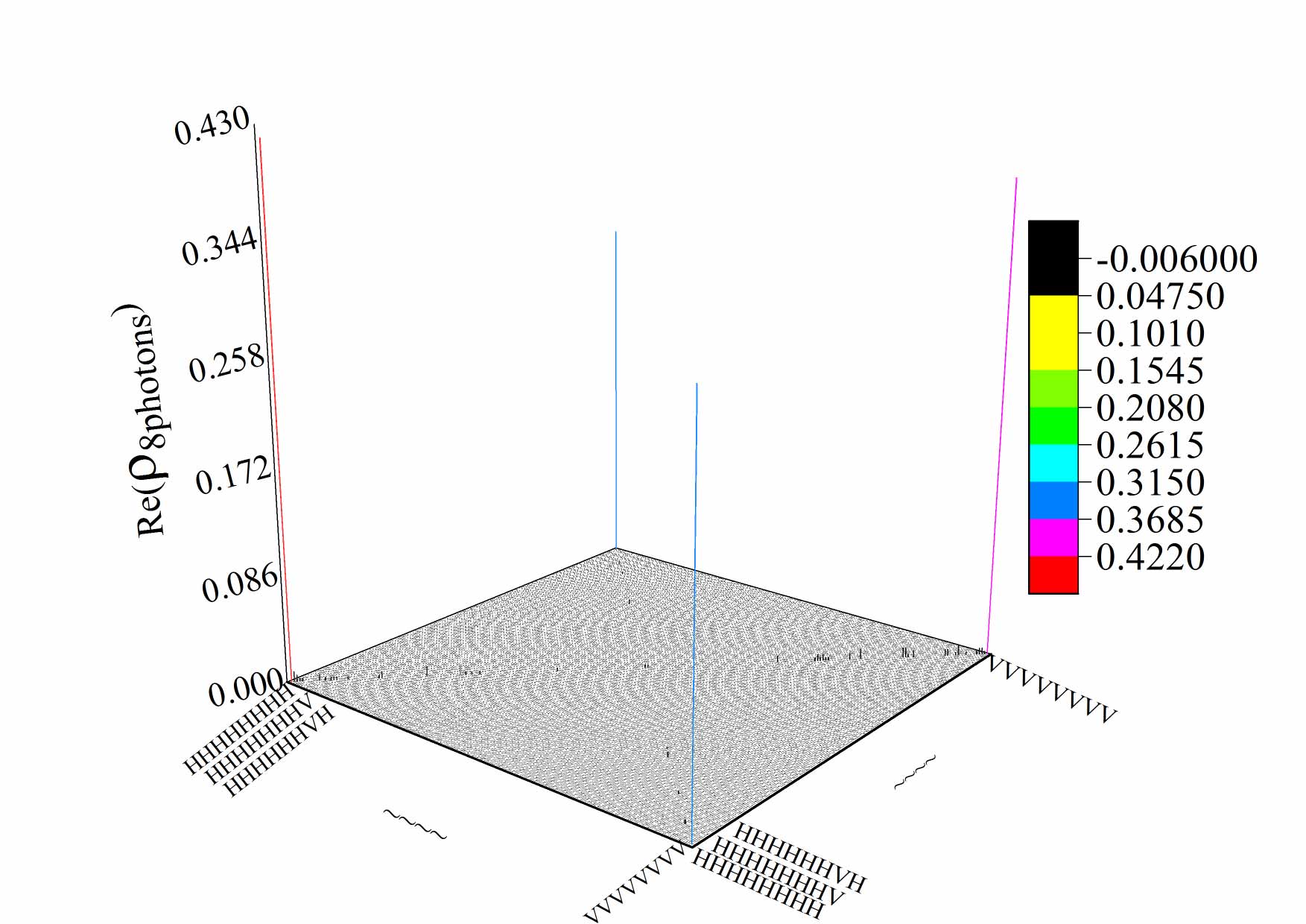}\\
  \caption{ The real part of experimental density matrix of eight-photon SC state ($\rho_{8photons}$). Different colors are applied to represent the values of the elements of density matrix.}\label{densitymatrix}
  \end{center}
\end{figure}

\begin{figure}[!h]
\begin{center}
  \includegraphics[width=10cm]{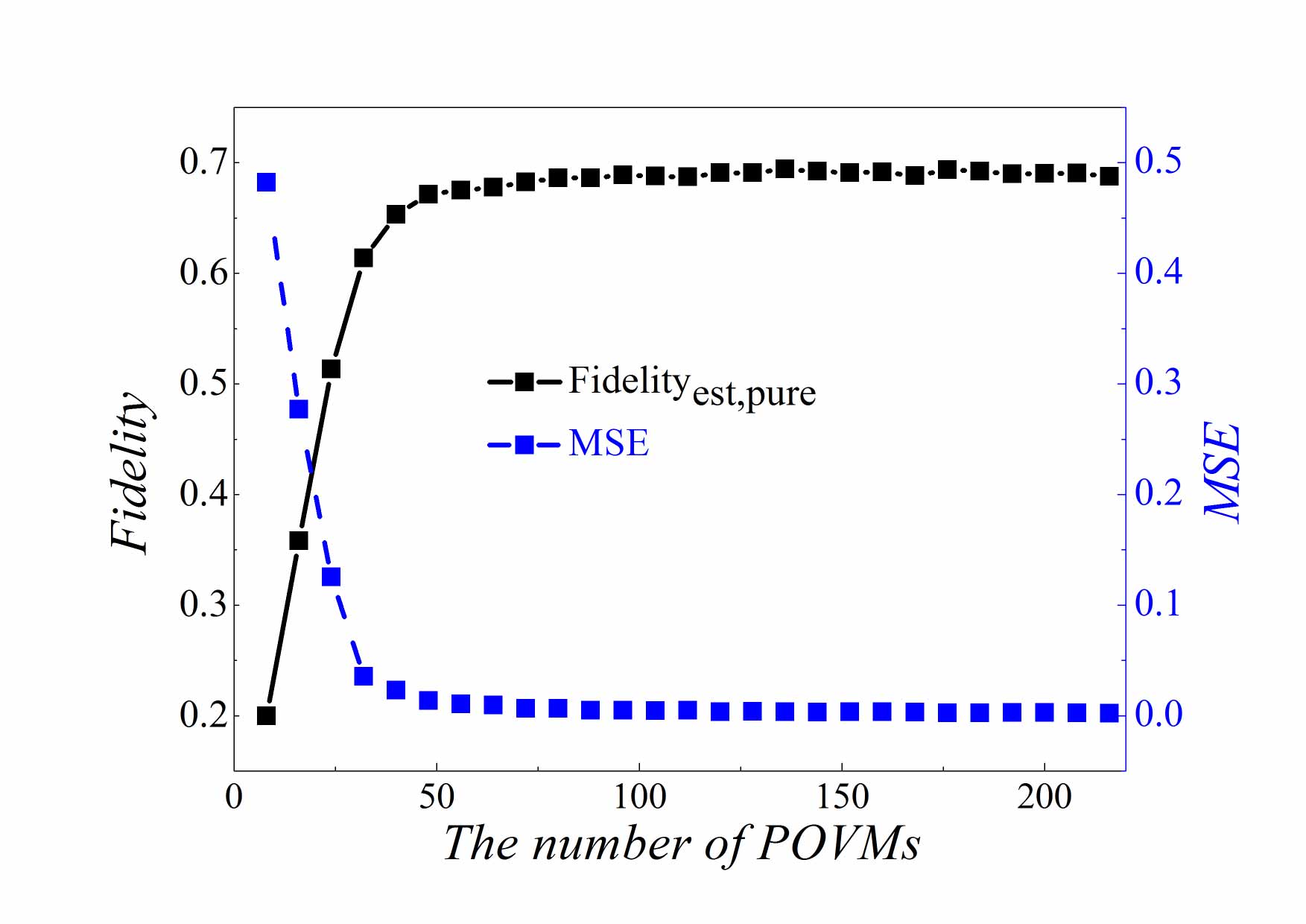}\\
  \caption{ Fidelities and MSEs' of state $\rho_{est}$ under different number of POVMs. $\rho_{est}$ is the estimated density matrix from $\rho_{exp_{-}3qubits}$. The fidelity is calculated by $Fidelity_{est,pure}=Tr(\rho_{est}|SC\rangle\langle SC|)$. The MSE (Mean square error) is calculated by $MSE=Tr[(\rho_{est}-\rho_{exp_{-}3qubits})^{\dagger}(\rho_{est}-\rho_{exp_{-}3qubits})]$. }\label{threefreprob}
  \end{center}
\end{figure}

 Several examples are given below in the following. The density matrix of the three-photon SC state ($\rho_{exp_{-}3qubits}$) is obtained by using the experimental data and construction method summarized by Eq.(\ref{optidensitymatirx}) with Pauli measurement, as shown in Figure \ref{Figure5789}a. The two large elements on the diagonal of the density matrix $\rho_{exp_{-}3qubits}$ are equal to 0.50188 for $|HHH\rangle\langle HHH|$ and 0.38419 for $|VVV\rangle\langle VVV|$. And the real parts of two main elements on the anti-diagonal are both 0.37238
on $|HHH\rangle\langle VVV|$ and $|VVV\rangle\langle HHH|$. The imaginary parts are quite small, so are not drawn. The density matrix of four-photon SC state ($\rho_{exp_{-}4qubits}$) is also obtained by using experimental data and phaselift, as shown in Figure \ref{Figure5789}b. The result is obtained by Pauli measurement and Eq.(\ref{optidensitymatirx}). Two large elements on the diagonal of the density matrix equal to 0.50637 for $|HHHH\rangle\langle HHHH|$ and 0.36161 for $|VVVV\rangle\langle VVVV|$. The real parts of elements on the anti-diagonal are 0.35944 on $|HHHH\rangle\langle VVVV|$ and $|VVVV\rangle\langle HHHH|$. Since the $\rho_{exp_{-}3qubits}$ (Figure \ref{Figure5789}a) and $\rho_{exp_{-}4qubits}$ (Figure \ref{Figure5789}b) have very small noise and the purity is high, density matrix of three-qubit ($\rho_{3qubits}$) (Figure \ref{Figure5789}c, Figure \ref{Figure5789}d) with much more noise is created for the following simulation. Two large elements on the diagonal of the density matrix are equal to 0.3716 for $|HHH\rangle\langle HHH|$ and 0.3412 for $|VVV\rangle\langle VVV|$. And the real parts of two main elements on the anti-diagonal are both 0.3504 on $|HHH\rangle\langle VVV|$ and $|VVV\rangle\langle HHH|$. The corresponding imaginary part is nearly approach to zero, so is not shown. The density matrix of the eight-photon system ($\rho_{8photons}$) is also drawn in Figure \ref{densitymatrix}. Obviously, only the real part of elements in four corners of the density matrix are larger than 0.2; other elements are much less than it, which is the characteristic of SC state. Besides, the imaginary part is too small; therefore, it is not drawn. By the density matrix of Figure \ref{Figure5789}a, we reconstructed the $\rho_{exp_{-}3qubits}$ under different number of pauli measurement. The reconstructed density matrix is $\rho_est$. Figure \ref{threefreprob} exhibits fidelities and Mean Square Error (MSE) when a different number of POVM is applied. When sampling number of POVM achieves around 45, fidelity is in a stable value (around 0.7) and the corresponding MSE is near 0.

\begin{figure}[!h]
\centering
\includegraphics[width=8cm]{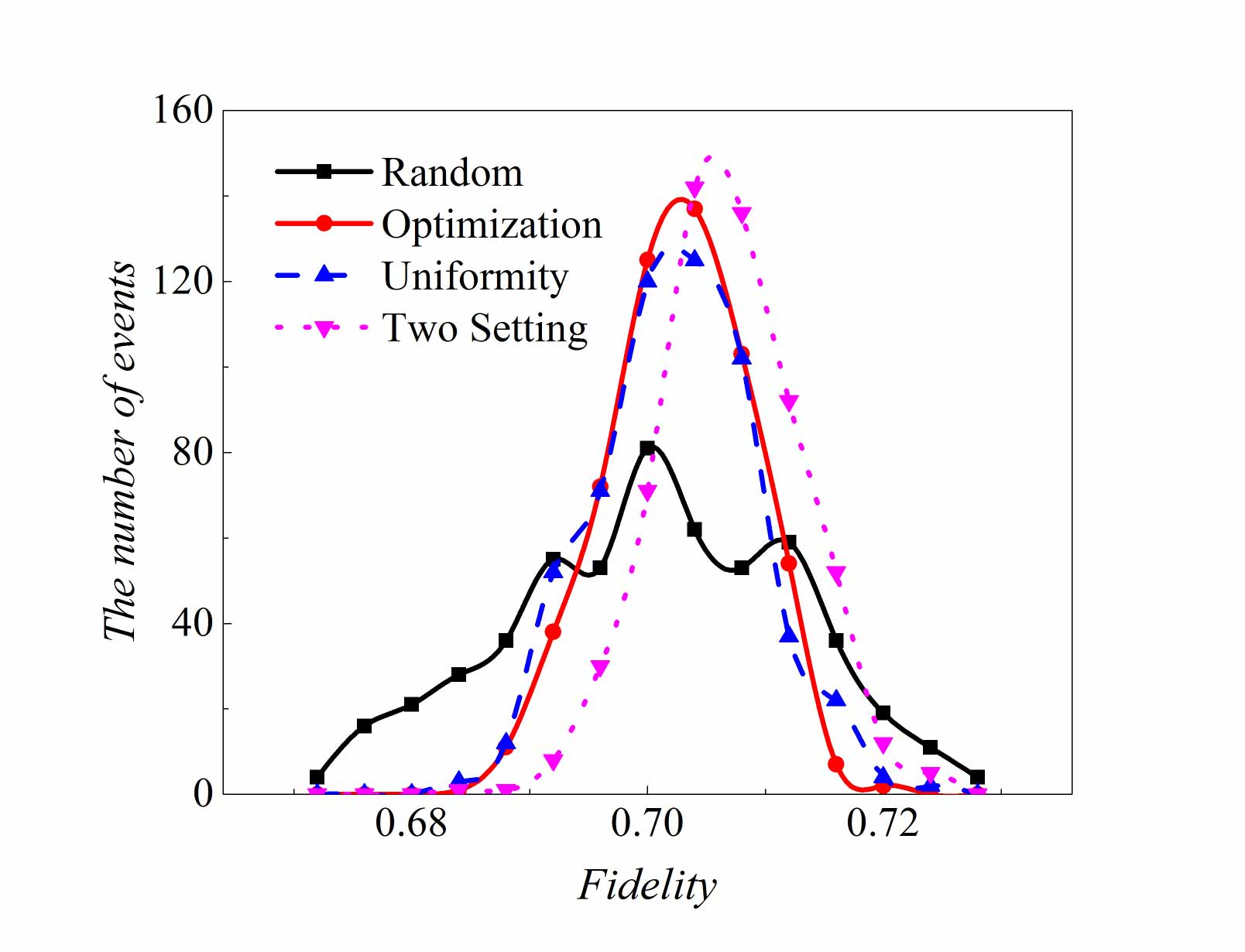}
\caption{ The distribution of fidelities between the target pure SC state and the estimated states under different number of copy distribution on different settings. $``Random"$ means the number of event of fidelity when distribution of copy of $\rho_{3qubits}$ goes $``500_{-}1000_{-}5000_{-}3500"$, in which the $500$ copies of state $\rho_{3qubits}$ are projected into $S_{3qubits,1}$;
$1000$ copies of $\rho_{3qubits}$ is projected into $S_{3qubits,2}$; $5000$ copies of $\rho_{3qubits}$ is projected into the basis of set of $S_{3qubits,3}$ and $3500$ copies of $\rho_{3qubits}$ is projected into $S_{3qubits,4}$. Similarly, ``Optimization" represents the event number of fidelity when distribution of copies of $\rho_{3qubits}$ is $``3630_{-}2570_{-}2670_{-}1130"$, in which $3630$ copies of state $\rho_{3qubits}$ are projected into $S_{3qubits,1}$; $2570$ copies of $\rho_{3qubits}$ is projected into $S_{3qubits,2}$; $2670$ copies of $\rho_{3qubits}$ is projected into $S_{3qubits,3}$ and 1130 copies of $\rho_{3qubits}$ is projected into $S_{3qubits,4}$. ``Uniformity" means the distribution is $``2500_{-}2500_{-}2500_{-}2500"$, which represents all is equal to $2500$ for the number of copies of state $\rho_{3qubits}$ that projected into $S_{3qubits,1}$, $S_{3qubits,2}$, $S_{3qubits,3}$ and $S_{3qubits,4}$. ``Two Setting" represents the both equals to 5000 for the copies of $\rho_{3qubits}$ that projected into $S_{3qubits,1}$ and $S_{3qubits,2}$. The range of fidelity is split into $250$ intervals on average between $0$ and $1$ to compare the event number. Different number of events that fidelity lies into a certain interval is gained, such as, if the calculated fidelity is $0.005$, then it belongs to the interval between $0.004$ and $0.008$, the number of events belong to this interval is added to $1$, and so forth. Fidelity is estimated for $550$ times in all four situations. Black squares represent the number of events accumulated in each interval for the case of $``500_{-}1000_{-}5000_{-}3500"$; Red circles represent for the case of $``3630_{-}2570_{-}2670_{-}1130"$; Blue triangles represent for the case of $``2500_{-}2500_{-}2500_{-}2500"$ and pink triangles represent for the case of $``5000_{-}5000"$. The fidelity between pure three-qubit SC state and $\rho_{3qubits}$ is $0.7068$. The points are connected by the lines with the same color of the points. It is observed that red circles and pink triangles have the most events near this value. Therefore both the ``Optimization" ($``3630_{-}2570_{-}2670_{-}1130"$) and ``Two setting" ($``5000_{-}5000"$), perform better for the estimation of fidelity. While the black squares for the random one gives the worst estimation.}\label{optimizeRandoAverage}
\end{figure}

\begin{figure}[!h]
\centering
\includegraphics[width=10cm]{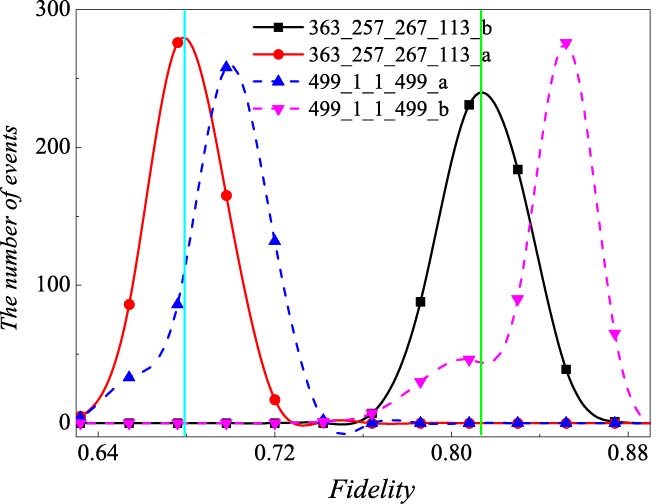}
\caption{The distributions of fidelities between target pure SC state or experiment $\rho_{exp_{-}3qubits}$ and estimations obtained from the 1000 copies of $\rho_{3qubits}$. The cyan vertical line at 0.7068 represents the fidelity between $\rho_{3qubits}$ and pure three-qubit SC state. Green vertical line at 0.82274 represents fidelity between experiment $\rho_{exp_{-}3qubits}$ and pure SC state. In the figure, ``a" represents the fidelity between the estimation and experiment $\rho_{exp_{-}3qubits}$. ``b" represents the fidelity between the estimation and target pure state. $``499_{-}1_{-}1_{-}499"$ represents the distribution of fidelity when $499$ copies of $\rho_{3qubits}$ are projected into the bases of $S_{3qubits,1}$; $1$ copy of $\rho_{3qubits}$ is projected into the bases of $S_{3qubits,2}$, $1$ copy of $\rho_{3qubits}$ is projected into $S_{3qubits,3}$ and $499$ copies of $\rho_{3qubits}$ are projected into $S_{3qubits,4}$. The number of events of fidelities between estimations and target pure state in this case is denoted by pink triangle, pink dashed line is applied to connect them. The number of events of fidelities between the estimation and experiment in this case is denoted by blue triangle and blue dashed line is applied to connect them. $``363_{-}257_{-}267_{-}113"$ represents fidelity distribution obtained from optimization distribution on four settings applied by 1000 copies of experimental three-qubit SC state ($\rho_{exp_{-}3qubits}$) built by Pauli measurements. $``363_{-}257_{-}267_{-}113"$ represents 363 copies of $\rho_{3qubits}$ are projected into $S_{3qubits,1}$, 257 copies of $\rho_{3qubits}$ is projected into $S_{3qubits,2}$, $267$ copies of $\rho_{3qubits}$ is projected into $S_{3qubits,3}$ and $113$ copies of $\rho_{3qubits}$ is projected into $S_{3qubits,4}$. Red circles are applied to denote the events of fidelities between the estimations and experiment. Red line is applied to connect them. Black squares represent the number of events of fidelity between the estimation and target pure state. Black line is applied to connect them. It is observed that $``363_{-}257_{-}267_{-}113_{-}a"$ performs better for fidelity estimation than $``499_{-}1_{-}1_{-}499_{-}a"$. $``363_{-}257_{-}267_{-}113_{-}b"$ performs better than $``499_{-}1_{-}1_{-}499_{-}b"$. Therefore, optimization distribution of copies of $\rho_{3qubits}$ performs better than the performance of two settings.
}\label{fifu1}
\end{figure}

Similarly, the following four measurement settings are defined for three qubits measurement.
\begin{eqnarray}
&&  S_{3qubits,1}=\{(|H\rangle\langle H|)^{\otimes3}, (|H\rangle\langle H|)^{\otimes2}(|V\rangle\langle V|),\cdots,(|V\rangle\langle V|)^{\otimes3}  \}, \nonumber\\
&&     S_{3qubits,2}=\{(|+,\theta\rangle\langle +,\theta|)^{\otimes3}, (|+,\theta\rangle\langle +,\theta|)^{\otimes2}(|-,\theta\rangle\langle -,\theta|),\cdots,\nonumber\\
&&(|-,\theta\rangle\langle -,\theta|)^{\otimes3}, \theta=0  \}, \nonumber\\
&&    S_{3qubits,3}=\{(|+,\theta\rangle\langle +,\theta|)^{\otimes3}, (|+,\theta\rangle\langle +,\theta|)^{\otimes2}(|-,\theta\rangle\langle -,\theta|),\cdots,\nonumber\\
&&(|-,\theta\rangle\langle -,\theta|)^{\otimes3}, \theta=\pi/3 \}, \nonumber\\
&&    S_{3qubits,4}=\{(|+,\theta\rangle\langle +,\theta|)^{\otimes3}, (|+,\theta\rangle\langle +,\theta|)^{\otimes2}(|-,\theta\rangle\langle -,\theta|),\cdots,\nonumber\\
&& (|-,\theta\rangle\langle -,\theta|)^{\otimes3}, \theta=2\pi/3 \}. \nonumber\\
\end{eqnarray}
Two measurement settings shown in \cite{55}, which is to project the state into $S_{3qubits,1}$ and $S_{3qubits,2}$, can also be applied to certify entanglement. For the three-qubit SC state, the density matrix ($\rho_{3qubits}$) are measured in four settings or two settings, respectively, and then the fidelity is estimated, as shown in Figure \ref{optimizeRandoAverage}. In the figure, it shows the distribution of number of event of fidelity between a target pure SC state and estimated one when 10000 copies of three-photon SC state $\rho_{3qubits}$ are measured. The fidelity between $\rho_{3qubits}$ and pure three-qubit SC state is 0.7068. We use $Random$ to represent the distribution of $500_{-}1000_{-}5000_{-}3500$, which represents the number of copies of $\rho_{3qubits}$ prepared in four settings, such as 500 is prepared in the
$S_{3qubits,1}$, 1000 is for the setting of $S_{3qubits,2}$, 5000 for $S_{3qubits,3}$, and 3500 for $S_{3qubits,4}$. Optimization means $3630_{-}2570_{-}2670_{-}1130$, in which 3630 copies of $\rho_{3qubits}$ is projected into the setting of $S_{3qubits,1}$, 2570 copies of $\rho_{3qubits}$ is projected into $S_{3qubits,2}$, 2670 is for the setting of $S_{3qubits,3}$, 1130 is for the setting of $S_{3qubits,4}$. $``Uniformity"$ means $2500_{-}2500_{-}2500_{-}2500$, which means all four settings are projected with the same number of copies of $\rho_{3qubits}$ (2500). ``Two setting" means
$5000_{-}5000$, which represents the number of copy of $\rho_{3qubits}$ prepared in two settings, such as 5000 is for $S_{3qubits,1}$, and 5000 for $S_{3qubits,2}$. Both two-setting distribution and the optimized distribution of copies of $\rho_{3qubits}$ ($3630_{-}2570_{-}2670_{-}1130$) gives the best estimation of fidelity, while the randomized distribution ($500_{-}1000_{-}5000_{-}3500$) gives the worst estimation. Christ mentioned that a bias exists for fidelity estimation when the semi-definite constraint is added to the maximum likelihood approach, and this bias is based on density matrix \cite{christ}. Here phaselift is applied, and there is no obvious bias for fidelity estimation when the number of copies of $\rho_{3qubits}$ approaches 10000, as shown in Figure \ref{optimizeRandoAverage}. However, there is an obvious bias when the number of copies of $\rho_{3qubits}$ drops to 1000 and the number for the setting of $S_{3qubits,2}$ is switched into the setting of $S_{3qubits,4}$ in two settings case, as exhibited in Figure \ref{fifu1}.

\begin{figure}[!h]
\centering
\includegraphics[width=10cm]{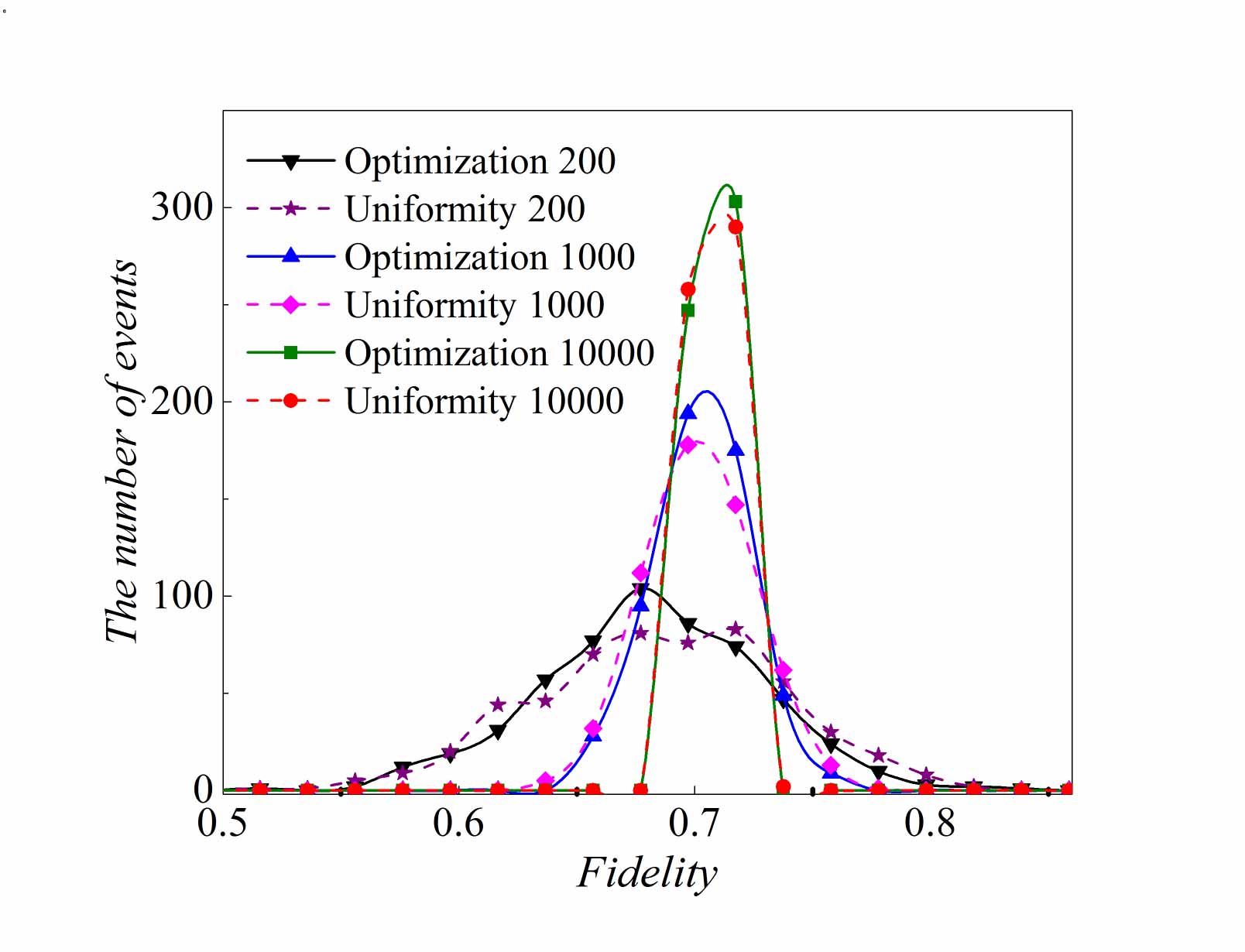}
 \caption{ The distributions of number for events of fidelities between estimations and target pure SC state when the fidelities are obtained from the density matrices that are constructed by 200, 1000 and 10000 copies of $\rho_{3qubits}$. The black down triangle connected by black solid line represents the number of events of fidelities when they are calculated by optimization distribution of total 200 copies of $\rho_{3qubits}$ on all the four settings. Purple star connected by dashed line represents the number of events of fidelities when it is calculated by uniform distribution of 200 copies of $\rho_{3qubits}$ on all the four settings. Blue up triangle connected by solid line represents the number of events of fidelities when it is calculated by optimization distribution of 1000 copies of $\rho_{3qubits}$ on all the four settings. Magenta diamond connected by dashed line represents the number of events of fidelities when it is calculated by uniform distribution of 1000 copies of $\rho_{3qubits}$ on all the four settings. Olive square connected by solid line represents the number of events of fidelities when it is calculated by optimization distribution of 10000 copies of $\rho_{3qubits}$ on all the four settings. Red circle connected by short dashed line represents the number of events of fidelities when it is calculated by uniform distribution of 10000 copies of $\rho_{3qubits}$ on all the four settings. Optimization distribution of 200 copies of $\rho_{3qubits}$ is $``73_{-}51_{-}53_{-}23"$, which represents the distribution of fidelity when $73$ copies of state $\rho_{3qubits}$ are projected into $S_{3qubits,1}$; $51$ copies of $\rho_{3qubits}$ is projected into $S_{3qubits,2}$, $53$ copies of $\rho_{3qubits}$ is projected into $S_{3qubits,3}$ and $23$ copies of $\rho_{3qubits}$ is projected into the bases of $S_{3qubits,4}$. Similarly, uniform distribution ($``50_{-}50_{-}50_{-}50"$), ($``250_{-}250_{-}250_{-}250"$), ($``2500_{-}2500_{-}2500_{-}2500"$), optimization distribution ( $``363_{-}257_{-}267_{-}113"$) and ($``3630_{-}2570_{-}2670_{-}1130"$) all follow the same rule as $``73_{-}51_{-}53_{-}23"$. Namely, the first number is the number of copy of $\rho_{3qubits}$ that projected into the setting of $S_{3qubits,1}$; the second number is the number of copy of $\rho_{3qubits}$ that projected into the $S_{3qubits,2}$; the third number is the number of copy of $\rho_{3qubits}$ that projected into $S_{3qubits,3}$ and the last number is the copies of $\rho_{3qubits}$ that is projected into the bases of $S_{3qubits,4}$. 10000 copies of $\rho_{3qubits}$ give much smaller error or standard deviation of fidelity than 200. Optimization always gives more centralized estimation of fidelity than uniform distribution under the same number of copies of $\rho_{3qubits}$.
 }\label{fi1}
\end{figure}

Fidelity estimation is also compared and analyzed in different initial conditions, such as the number of copies of $\rho_{3qubits}$, as shown in Figure \ref{fi1}, which presents that 10000 copies of $\rho_{3qubits}$ provide much better estimation of fidelity than 200 copies of the state. Furthermore, optimization always gives better estimation of fidelity than uniform distribution of copies of $\rho_{3qubits}$.

\section*{Optimization of the number of copies via experimental feedback}

\begin{figure}[!h]
\centering
\includegraphics[width=12cm]{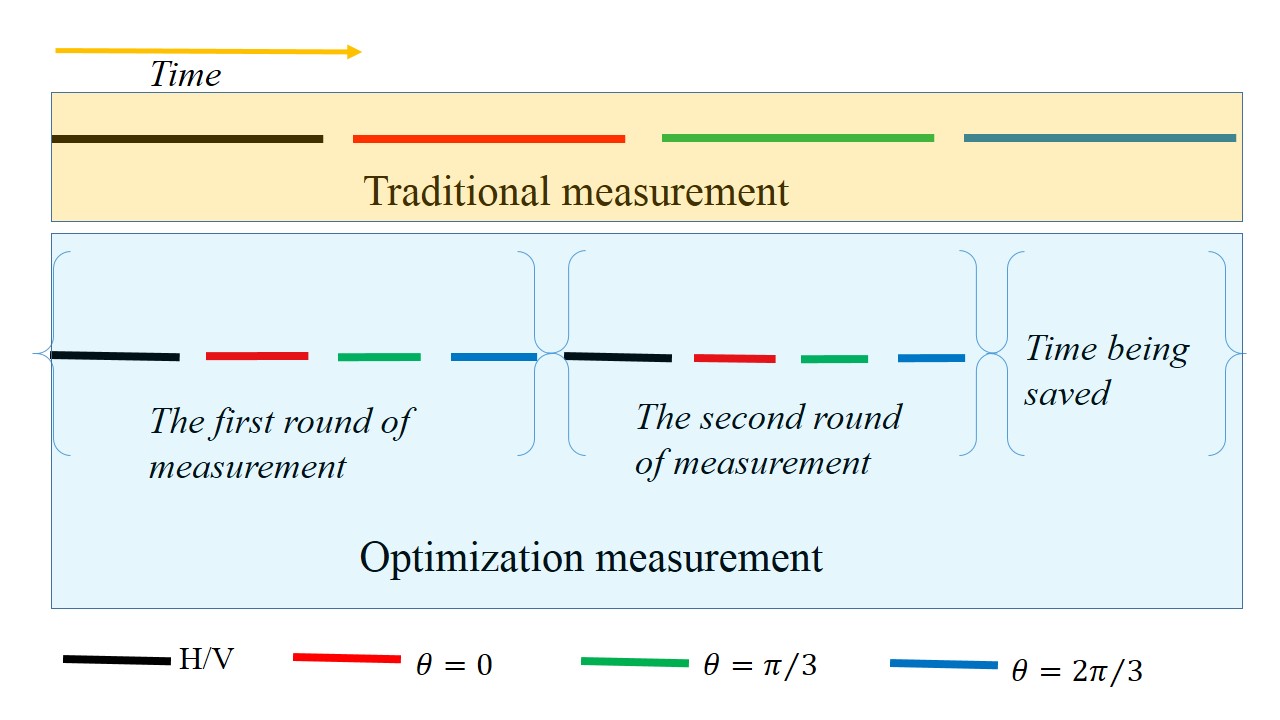}
\caption{ Comparison between traditional and optimization measurement of three-qubit state. The color of line segment represents the different setting. The length of line segment represents the time for the corresponding measurement. Traditional measurement order is to finish the measurement of each setting one by one, as shown by the Pastel yellow area. The optimization measurement is iterated twice as shown by the light blue area.}\label{celiangfangxiang}
\end{figure}

\begin{figure}[!h]
\centering
\includegraphics[width=10cm]{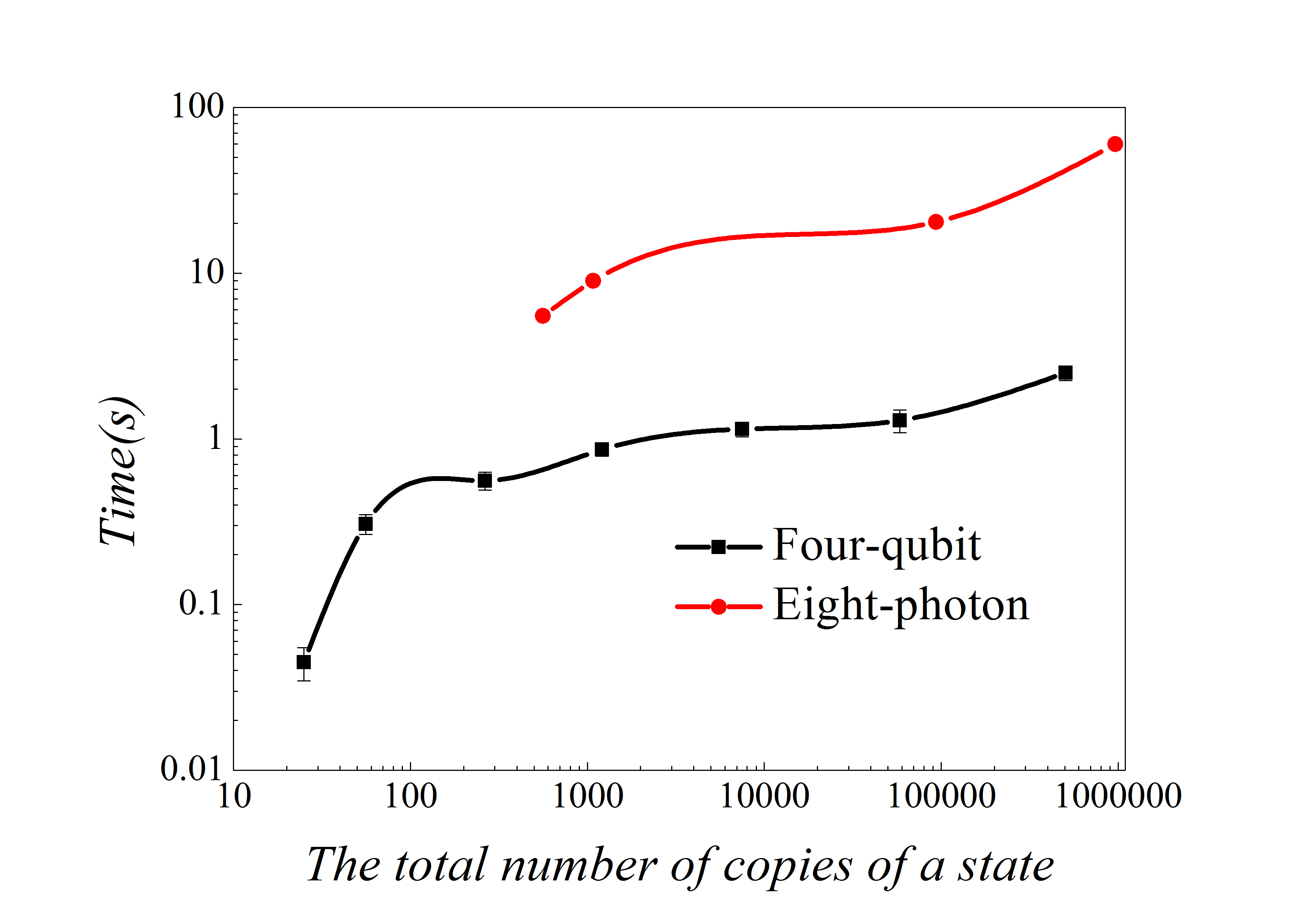}
\caption{\label{RunTimeVSCopyNumber} The required time of optimization for different number of copies of a state. The red circle represents running time of eight-photon optimization. Black square represents running time of four-qubit optimization. It is observed that the total time to calculate Eq.(\ref{optimalt}) and simulate the experiment is less than 100 seconds for both cases. Therefore compared with several hours spend to prepare copies of eight-photon state in experiment. It can be negelected.}
\end{figure}

Let us note that $P_j$ in Eq.(\ref{optimalt}) should be known before calculating $t_j$, and there is no way to obtain the precise value of $P_j$ without knowing density matrix via Born's rule or without experimental measurement. In ``Direct estimation of fidelity for experimental eight-photon SC state and simulated ten-photon SC state", we prior estimate a density matrix based on the preparation of copies experimentally. In ``Optimization of multi-qubit experimental and simulated data via density matrices", we estimate a precise density matrix by phaselift. Here, we show to calculate them through the experiment itself. Take eight-photon SC state experiment as an example. In experiment, pure eight-photon SC state is the target state that needs to be prepared. It can be taken as a priori to approximately decide $P_j$, such as $P_1$ is a value near to ${\rm Tr}((|H\rangle\langle H|)^{\otimes 8} |SC\rangle\langle SC|)+{\rm Tr}((|V\rangle\langle V|)^{\otimes 8} |SC\rangle\langle SC|)$, $P_j$ is near to a value given by Eq.(\ref{p1pj1pj}), $j=2, 3, \cdots, n+1$. However, the experimentally prepared state is not pure $|SC\rangle$ and it takes too long time to precisely estimate $P_j$, an optimization procedure is proposed based upon the experiment. It divides the process of measurement into a few steps. Instead of measuring one setting for a prolonged time to obtain the frequencies within small error margins, and then continue to measure the next setting for the same time, and so on. We divided this total long time into several intervals, and changed the order of measurements. The order is to measure all the required settings one by one for a much shorter time, then based on the measurement results; the $P_j$ can be estimated roughly. After that, the extra number of copies of a quantum state that needs to be prepared and measured for each setting can be given by Eq.(\ref{optimalt}) by inputting the rough $P_j$.
Afterwards, more copies of the quantum state are prepared and measured according to the $t_j$ given. Later, more precise frequencies can be
obtained and this process can be repeated until the final precision for fidelity is reached. Figure \ref{celiangfangxiang} shows the measurement order when the process is conducted only twice. We simulated this process in computer, it only costs a very short time, as shown in Figure \ref{RunTimeVSCopyNumber}.

To be specific, the main process is as following. Firstly, we introduce a superscript to represent the number of steps in optimization. The superscript $l$ of a parameter represents the parameter applied in the $l$-th step, i.e. $\epsilon^{1}$ represents the value $\epsilon$ used in Eq.(\ref{optimalt}) for the first round of measurement. At the beginning of fidelity estimation, $\epsilon^{1}$ is set to a large number, such as 0.01, and all of $P_j$ are originally set to $P_j=P_j^1=1/2$, $j=1,2,3,\cdots,n+1$, ($P_j^1$ can also be chosen according to target pure state $|SC\rangle$, such as $P_1^1$ can be a value near ${\rm Tr}((|H\rangle\langle H|)^{\otimes n} |SC\rangle\langle SC|)+{\rm Tr}((|V\rangle\langle V|)^{\otimes n} |SC\rangle\langle SC|)$ so that the suitable solution $t_j^l$ can be obtained by solving Eq.(\ref{optimalt}). The experiment is performed according to the $t_j^1$ copies of the quantum state. When all $t_j^1$ copies of the prepared quantum state is projected into a measurement setting, $P_j^2$ can be obtained. After that, input $P_j^2$ instead of $P_j^1$, and have $\epsilon^1$ become smaller; consequently, the $t_j^2$ can be obtained. Then copies of the state with the number of $t_j^2-t_j^1$ are projected into the $j$th measurement setting in the second round of experiment, so on and so forth. Measurement is ended when $\epsilon^{iter}$ is sufficiently small, obvious,
\begin{eqnarray}
\epsilon^{1}>\epsilon^{2}>\epsilon^{3}>\cdots>\epsilon^{l}\cdots>\epsilon^{iter}.
\end{eqnarray}

Extra time is needed for optimization; however it is much shorter compared with the time required for the preparation of copies of multi-photon entanglement state, as shown in Figure \ref{RunTimeVSCopyNumber}. The iteration makes the experiment to have more pauses between different settings during the measurement process. Mostly, switching settings cost more time, generally is about 3 or 4 times of the switching time in conventional measurement. Anyhow the time required for optimization is much shorter than that spent on the preparation of the copies of multi-photon entanglement state. Generally, switches and optimizations only cost less than two minutes, while the coincidence count rate of eight-photon entanglement state is too low that it costs several hours to produce just enough copies of the state for only one setting. Therefore, the time in calculation and switching time can be neglected compared to the preparation of copies of multi-photon entanglement state.

In the following, a specific example is given. Numerical simulation applies a four-qubit SC state mixed with gaussian noise. The density matrix is $\rho_{4qubits}$, which fulfills trace equal to one and semi-definite condition. The fidelity between $\rho_{4qubits}$ and pure SC state is 0.9374. In simulation, parameters are chosen as $\epsilon^{1}=0.01$, $\epsilon^{2}=0.001$, $\epsilon^{3}=0.0001$, $\epsilon^{4}=0.00001$. The five measurement settings are required and listed as follows:
\begin{eqnarray}
&&  S_{4qubits,1}=\{(|H\rangle\langle H|)^{\otimes4}, (|H\rangle\langle H|)^{\otimes3}(|V\rangle\langle V|),\cdots,(|V\rangle\langle V|)^{\otimes4}  \}, \nonumber\\
&&  S_{4qubits,2}=\{(|+,\theta\rangle\langle +,\theta|)^{\otimes4}, (|+,\theta\rangle\langle +,\theta|)^{\otimes3}(|-,\theta\rangle\langle -,\theta|),\cdots,\nonumber\\
&& (|-,\theta\rangle\langle -,\theta|)^{\otimes4}, \theta=\pi/4  \}, \nonumber\\
&&  S_{4qubits,3}=\{(|+,\theta\rangle\langle +,\theta|)^{\otimes4}, (|+,\theta\rangle\langle +,\theta|)^{\otimes3}(|-,\theta\rangle\langle -,\theta|),\cdots,\nonumber\\
&& (|-,\theta\rangle\langle -,\theta|)^{\otimes4}, \theta=2\pi/4 \}, \nonumber\\
&&  S_{4qubits,4}=\{(|+,\theta\rangle\langle +,\theta|)^{\otimes4}, (|+,\theta\rangle\langle +,\theta|)^{\otimes3}(|-,\theta\rangle\langle -,\theta|),\cdots,\nonumber\\
&&(|-,\theta\rangle\langle -,\theta|)^{\otimes4}, \theta=3\pi/4 \}, \nonumber\\
&&  S_{4qubits,5}=\{(|+,\theta\rangle\langle +,\theta|)^{\otimes4}, (|+,\theta\rangle\langle +,\theta|)^{\otimes3}(|-,\theta\rangle\langle -,\theta|),\cdots,\nonumber\\
&& (|-,\theta\rangle\langle -,\theta|)^{\otimes4}, \theta=\pi \} \nonumber\\
 \end{eqnarray}

\begin{figure}[!h]
\centering
\includegraphics[width=10cm]{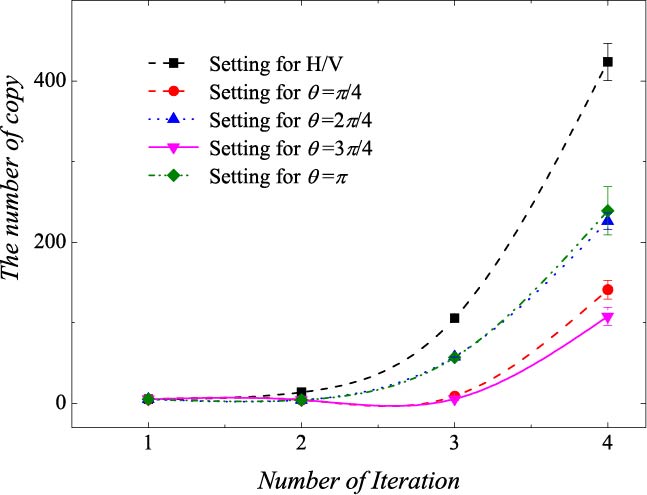}
\caption{\label{copieoptimiz}  The change of optimization number of copy of $\rho_{4qubits}$ of fidelity estimation corresponding to different iteration numbers with different settings: $\pi/4$, $2\pi/4$, $3\pi/4$, and $\pi$. ``Setting for H/V" represents the number of copy of $\rho_{4qubits}$ projected into $S_{4qubits,1}$. ``Setting for $\theta=\pi/4$" represents for $S_{4qubits,2}$, ``Setting for $\theta=2\pi/4$" represents for $S_{4qubits,3}$, ``Setting for $\theta=3\pi/4$" represents for $S_{4qubits,4}$, ``Setting for $\theta=\pi$" represents for $S_{4qubits,5}$. Error bar represents one standard deviation.}
\end{figure}

 All the initial numbers of copies of $\rho_{4qubits}$ for each setting are set at 5. Other initial parameters for 5 measurement settings are set at: $P_1=1/2$, $P_2=1/2$, $P_3=1/2$, $P_4=1/2$, $P_5=1/2$, respectively.
 In Figure \ref{copieoptimiz}, the $\rho_{4qubits}$ is taken as the test matrix. Its horizontal axis represents the number of iteration, which means the number of round of measurement. The corresponding point is the average number of extra copies of the state $\rho_{4qubits}$ that needs to be projected into each setting for the next round of measurement. The curve connects the number of required copies of $\rho_{4qubits}$ for each same setting. The error bar is one standard deviation, which is obtained by repeating the optimization program for 100 times. When the iteration ends, the $P_j$ is listed in Table \ref{lastpjandkj}.

\begin{table}[h]
\begin{center}
\begin{tabular}{|c|c|c|c|}
  \hline
$P_1$ &0.9339	\\
$P_2$&0.0316	\\
$P_3$& 0.9491	\\
$P_4$& 0.0325	\\
$P_5$&  0.9474  \\
  \hline
\end{tabular}
\caption{\label{lastpjandkj} The final $P_j$ when iteration ends.}
\end{center}
\end{table}

\begin{figure}[!h]
\begin{center}
  \includegraphics[width=10cm]{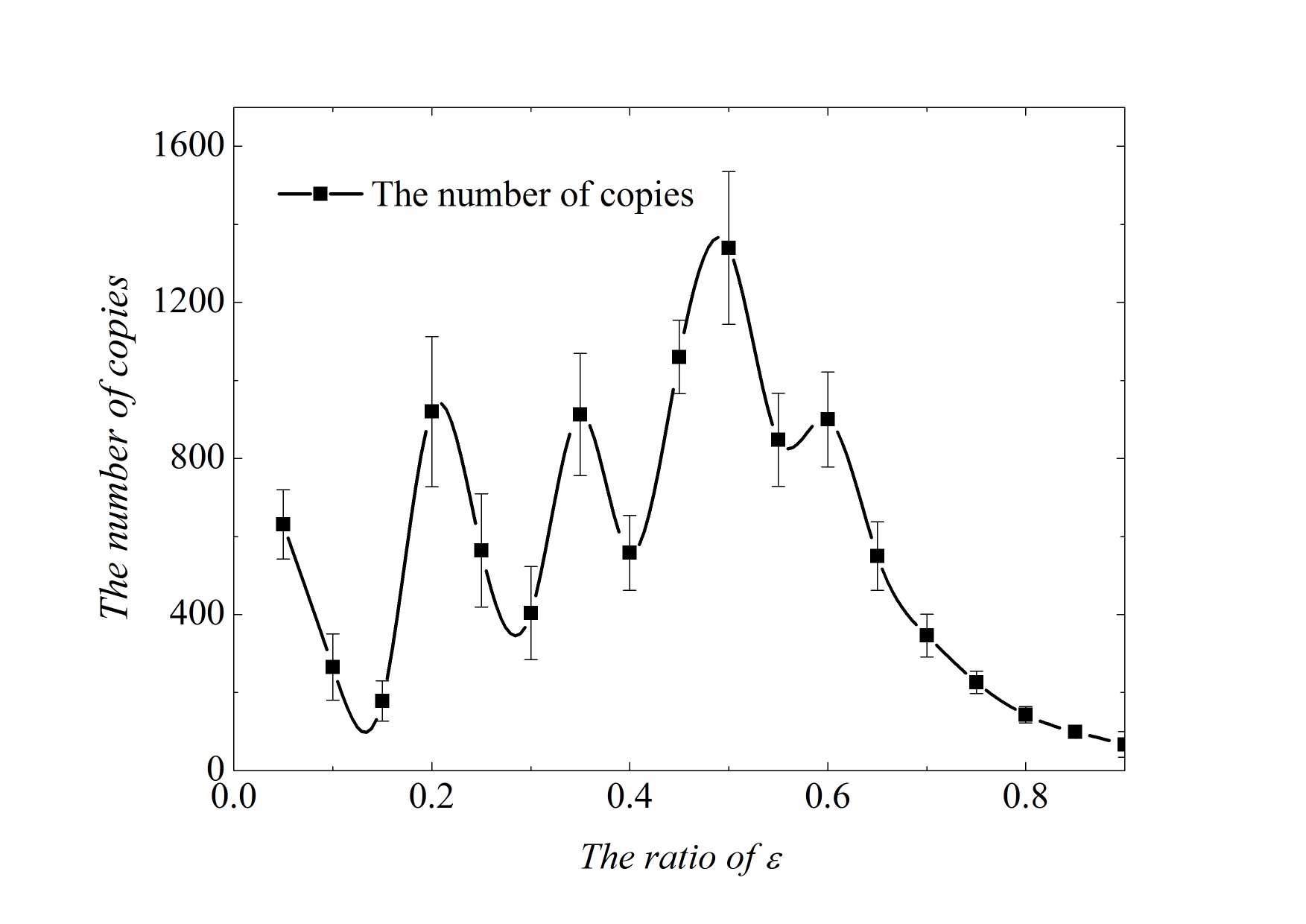}\\
  \caption{ The required copies of a state for different ratio of $\epsilon$. For each ratio of $\epsilon$, numerical test is conducted for 10 times. Black square is used to represent the number of copies of the state. Error bar is mean standard deviation. It is observed that the number of copies of state rises at wave type when the ratio of $\epsilon$ increases but no larger than 1/2 and decreases with the ratio of $\epsilon$ when it is larger than 1/2.}\label{copiep}
  \end{center}
\end{figure}

  We define the ratio between the $\epsilon^l$ applied in the current round of measurement and the $\epsilon^{l-1}$ applied in the previous round of measurement as the ratio of $\epsilon$, and let the ratio of $\epsilon$ for different round of measurement be equal to each other, that is $\epsilon^{2}/\epsilon^{1}=\epsilon^{3}/\epsilon^{2}=\cdots=\epsilon^{l}/\epsilon^{l-1}=\cdots$. The different values are tested to search for the best ratio that costs the least number of copies of a state. Figure \ref{copiep} shows the number of copies of state randomly created at different ratios of $\epsilon$ when it equals to a value between 0.05 and 0.9. It is observed that it requires most number of copies when the ratio is 1/2, it increases at a wave type before this value, and decreases after this value. Specifically, the following procedure is conducted. Initially, the number of copies of a randomly created state is an integer between $4$ to $7$ for each setting, $P_1$, $P_2$, $\cdots$, $P_5$ are all given a value randomly created between 0.25 and 0.75 and $\epsilon$ is $0.01$. Then, a fixed ratio of $\epsilon$, such as 0.1, is applied. It means $\epsilon$ is set to $0.001$ in the second round of measurement, $0.0001$ in the third round of measurement and so on. Eq.(\ref{optimalt}) is applied to calculate the needed copies ($t_j$) of the state for each round of measurement. Then $t_j$ is summed up to gain the total number of the copies of the state in current round of measurement ($\sum_jt_j$). The iteration stops when $\epsilon$ is smaller than 0.0003. The minimum number of copies of the state can be found by repeating the above steps by changing the ratio of $\epsilon$. It is found that the value near 0.15 for the ratio requires the least number of copies of a randomly created state, which is 178. The total iteration number for each setting for most created states is about 3 or 4 to achieve the 0.0003 of finial $\epsilon$. It is also noticed that the required number of copies is even less when the ratio of $\epsilon$ approaches 0.9. However, it is not suitable to apply the large value for the ratio of $\epsilon$ since too less copies of a state may lead to our model hold with a low probability, which will be discussed in the following section.

\section*{Discussion}
\subsection*{Characteristics of optimization of the successful probabilities}\label{ivc}

It is noticed that the summation of number of copies of an unknown state projected into the same setting must be larger than a certain value, since a large value can confirm the model to hold with probability nearly one.

In the above sections, the minimum number of copies of an unknown state is obtained and its fidelity belongs to the interval with the certain high probability. Hoeffding's inequality is a mathematical way to describe the probability. It states that the sample average $\overline{X}=\sum X/t$ of $t$ independent, not essentially identical distributed, bounded random variables with $Prob[X_i\in[a_i,b_i]]=1$ for $i=1,2,\cdots,t$ satisfies
\begin{eqnarray}
&& Prob\left[\overline{X}-E[\overline{X}]\leq-h\right]\nonumber\\
&&\leq exp[-2t^2h^2/\sum(b_i-a_i)^2]
\end{eqnarray}
for all $h>0$, where $X_i$ is a variable, $a_i$ is the lower bound, $b_i$ is the upper bound, $t$ is the number of samples, $E[\overline{X}]$ denoting the mean value of $\overline{X}$, $h$ is the definite value that equals to the maximum deviation from expectation \cite{Hoeffdingone, Hoeffdingtwo}.

Now this inequality is applied to multi-photon entanglement certificate experiment. The measurement response of a single copy of an unknown state is taken as the value of a single random variable. Since photon detectors can only give the feedback, $0$ or $1$, this leads to $b_i=1$ and $a_i=0$. $\overline{X}$ corresponds to a relative frequency, which is denoted to be $f_j$, $j$ is to distinguish different measurement settings. Hence the expectation $E[\overline{X}]$ corresponds to the probability $p_j$. The total copy of a state for the $j_{th}$ setting is represented by $t_j$ instead of $t$. By replacing all of them, we obtain
\begin{eqnarray}
&& Prob\left[f_j-p_j\leq-h_j\right]\leq exp[-2t_jh_j^2]\label{probb},
\end{eqnarray}
where $h_j$ is the deviation from true probability $p_j$. It means
\begin{eqnarray}
&& Prob\left[f_j\in (p_j-h_j,p_j+h_j)\right]\geqslant 1-2exp[-2t_jh_j^2].
\end{eqnarray}
Then
\begin{eqnarray}
&& Prob\left[1-f_j\in (1-p_j-h_j,1-p_j+h_j)\right]\nonumber\\
&& \geqslant 1-2exp[-2t_jh_j^2].
\end{eqnarray}
Therefore,
\begin{eqnarray}
&& Prob\left[f_j-(1-f_j)\in (2(p_j-h_j)-1,2(p_j+h_j)-1)\right]\nonumber\\
&& \geqslant 1-2exp[-2t_jh_j^2].
\end{eqnarray}
Let $f_j-(1-f_j)$ be $p_j$. Hence,
\begin{eqnarray}
&& Prob\left[P_j\in (2(p_j-h_j)-1,2(p_j+h_j)-1)\right]\nonumber\\
&&\geqslant 1-2exp[-2t_jh_j^2],j=2,3,\cdots,n+1.
\end{eqnarray}
Let $2(p_j-h_j)-1$ be $P_j^-$ and $2(p_j+h_j)-1$ be $P_j^+$, and based on Eq. (\ref{k1j}), $k_j\in[k_j^-,k_j^+]$,
where
\begin{eqnarray}
&& k_1^+=Max{\left\{\frac{1}{4}P_1^+(1-P_1^+), \frac{1}{4}P_1^-(1-P_1^-)\right\}},
\end{eqnarray}
\begin{eqnarray}
&& k_1^-=Min{\left\{\frac{1}{4}P_1^+(1-P_1^+), \frac{1}{4}P_1^-(1-P_1^-)\right\}},
\end{eqnarray}
\begin{eqnarray}
&& k_j^+=Max{\left\{\frac{1}{n^2}P_j^+(1-P_j^+), \frac{1}{n^2}P_j^-(1-P_j^-)\right\}},\nonumber\\
&& j=2, 3, \cdots, n+1,
\end{eqnarray}
\begin{eqnarray}
&& k_j^-=Min{\left\{\frac{1}{n^2}P_j^+(1-P_j^+), \frac{1}{n^2}P_j^-(1-P_j^-)\right\}}, \nonumber\\
&&j=2, 3, \cdots, n+1,
\end{eqnarray}
then from Eq.(\ref{optimalt}), one has
\begin{eqnarray}
&& t_i^+=\frac{\sqrt{k_i^+}\left(\sum_{j=1}^{n+1}\sqrt{k_j^+}\right)}{\epsilon},
\end{eqnarray}
\begin{eqnarray}
&& t_i^-=\frac{\sqrt{k_i^-}\left(\sum_{j=1}^{n+1}\sqrt{k_j^-}\right)}{\epsilon}.
\end{eqnarray}

Therefore the $t_i$ $\in[t_i^-,t_i^+]$ in Eq.(\ref{optimalt}) when the holding probability of model Eq.(\ref{Optimizationproblem}) is considered.

 Obviously, $h_j$ has the impact on $t_i^+$ and $t_i^-$. The larger $h_j$ is, the larger the gap between $t_i^+$ and $t_i^-$. Large $h_j$ and $t_i$ from Eq.(\ref{probb}) are needed to keep results with high probability. However, large $t_i$ costs too much experimental time. Large $h_j$ may introduce too large a gap between $t_i^+$ and $t_i^-$, which may then lead to the wrong number of copies of an unknown state. Therefore, it requires to choose suitable $h_j$ and $t_i$.

 By comparing the optimization results with the experiment, it is found that only $986$ copies of $\rho_{8qubits}$ are used compared with the $1305$ copies of $\rho_{8qubits}$ in eight photon experiment, which specifies that $24$ percent copies of eight-photon SC state ($\rho_{8qubits}$) can be saved. Specifically, $h_j$ is chosen to be $0.2$ for all $j$. According to joint probability, $\prod_jp_j$ is calculated, in which $p_j$ is the successful probability for each setting. For eight photon measurement, $j=1,2,\cdots,9$, the final probability is 0.9972 for experiment after $1305$ copies of $\rho_{8qubits}$ are measured. We observed same probability is obtained when $110$ copies of $\rho_{8qubits}$ for each setting are used and all $h_j$s' are chosen as 0.2.

\begin{figure}[!h]
\centering
\includegraphics[width=10cm]{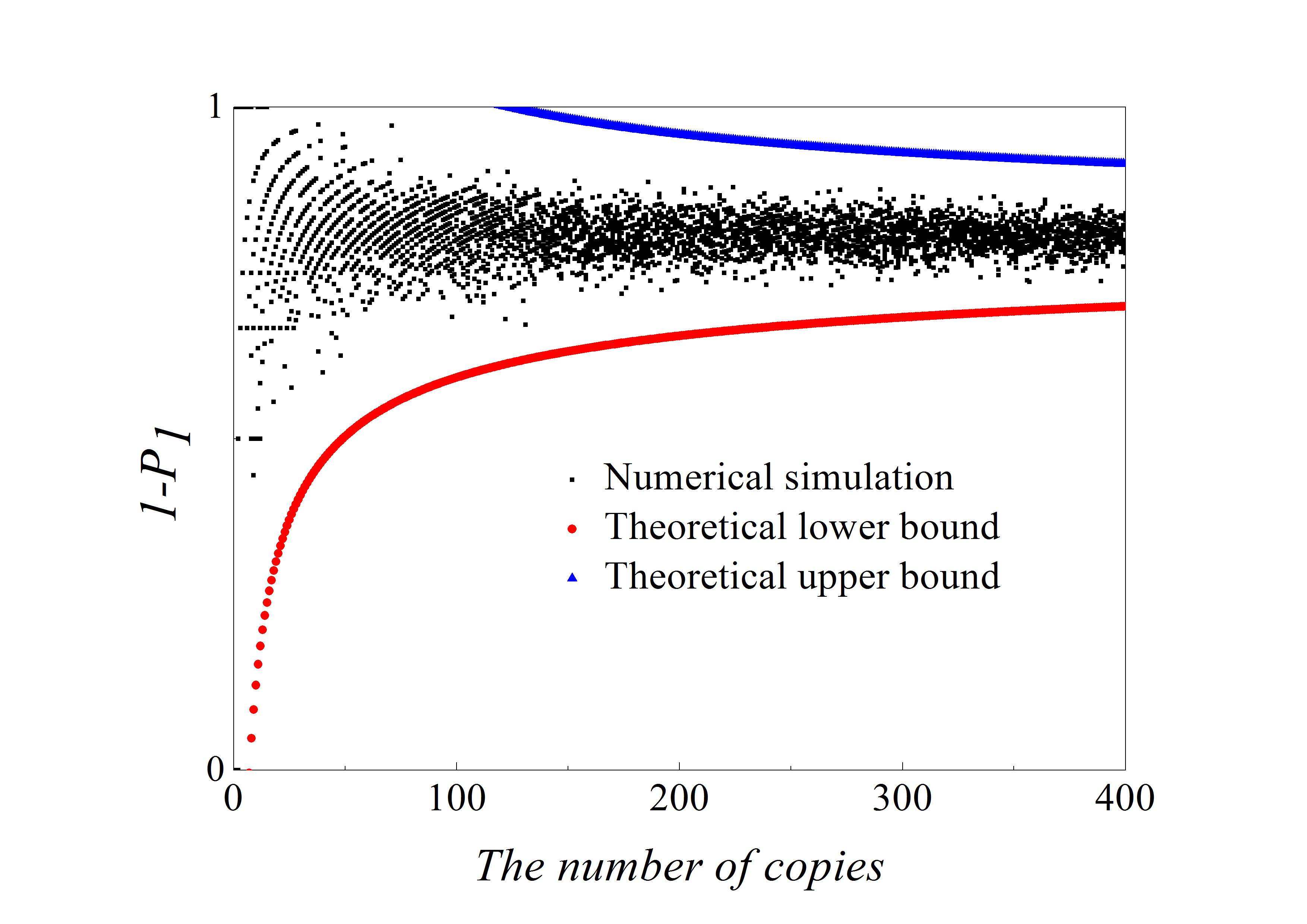}
\caption{\label{probaibcopy} The change of $1-P_1$ in the setting of $S_1$ corresponding to different number of copies of eight-photon SC state($\rho_{_8qubits}$). Black square represents the numerical simulation, red circle represents the theoretical lower bound and upper triangle represents the theoretical upper bound. Numerical simulation is repeated for 10 times for each number of copies. It is observed that all the simulated points lie in the region that consists of point that is larger than the lower bound and smaller than the upper bound.}
\end{figure}

In the above analysis, we assume Hoeffding's inequality describes the probability precisely. In the following, numerical simulation is produced to confirm the above mathematical tool is true. The density matrix ($\rho_{8qubits}$) is calculated from experimental frequencies, and new relative frequencies are obtained under a certain number of copies of $\rho_{8qubits}$ in a random simulation of experimental process that gets the relative frequency by computer. $1-P_1$ is the summation of relative frequency that all the qubits projected into horizontal polarization and the relative frequency that all the qubits projected into vertical polarization. The real value of $1-P_1$ is $0.8068$ when the number of copy is sufficiently large. When failing probability is set less than $0.0001$, Figure \ref{probaibcopy} shows how $1-P_1$ behaves under different number of copy of $\rho_{8qubits}$. In the figure, red circle and blue triangle are drawn according to Hoeffding's inequality, $1-P_1$ can be estimated much more precisely with an increasement of prepared copies of the state $\rho_{8qubits}$. It is observed that all the numerical simulated points lie in the region between upper bound and lower bound. Therefore, Hoeffding's inequality can be applied to describe the $P_j$ in multi-photon entanglement.


\section*{Extension of the optimization of the number of copies of a state to quantum-state tomography}

 The surprising thing brought to us by the optimization in the fidelity estimation, is to extend it to quantum state tomography. The optimization model for tomography is constructed as follows: Let $\rho_0$ be a $d\times d$ density matrix of real experimental created and $\rho$ be the estimated density matrix via limited copies of $\rho_0$.
For $n$ qubit state tomography, we build
\begin{eqnarray}
&&{\rm Minimize}\ \sum_{\nu=1}^{n_s}T_{\nu}\  \nonumber\\
&&{\rm subject\ to}\  \ ||\rho-\rho_0||_F\leq \epsilon_0,\label{Optimizationtomography}
\end{eqnarray}
where $T_{\nu}$ represents the number of copies of $\rho_0$ of the $\nu$th setting, $\nu$ can distinguish different measurement settings, $n_s$ is total number of measurement setting. The solution of Eq.(\ref{Optimizationtomography}) is

See ``Theoretical derivation of minimum number of copies of a state in quantum-state tomography" for the details to get the solution of Eq.(\ref{Optimizationtomography})

\section*{Conclusions}
We proposed an optimal approach that assists to find the minimum distribution of copies of a state that is sufficient to certify the entanglement of the state by fidelity. The main purpose is to facilitate an experiment to obtain better measurement strategy for fidelity estimations, for example, by changing the ratio of the number of copies of the state in different settings. To estimate fidelity directly from fewer copies of SC state (1253 copies), with optimized distribution, almost the same distribution of fidelity as the experimental one (1305) can be obtained. It not only saves time, about five percent of measurement time (6 hours) is saved, but also small error of fidelity. Additionally, the distribution on the number of copies of ten-photon SC state is also simulated, and 22.45\% of copies of the ten-photon SC state are saved, which further highlights the superiority of this scheme, and reveals that the optimized distribution of copies of a state in different settings gives better estimation of the fidelity than uniform distribution of copies of a state in all settings. Fidelity can also be estimated by the reconstructed density matrix. It is observed that the optimized distribution provides the best estimation of the true state, the uniform distribution provides a worse estimation, while randomized distribution provides the worst estimation. With the increase of the number of copies of the state the differences between different distributions (uniform distribution and optimized distribution) become much smaller. Besides the state with high similarity with SC state, this approach can also be extended into other states in parallel. And the scheme is extendable to tomography when the MSE between the estimated density matrix and real density matrix is limited to a fixed value.


\appendix

\section*{Preparation of ten-photon SC state}

From second paragraph of Reference \cite{2}, the count rate of eight-photon event is about $2.8\times10^{-5}Hz$. Accidental coincidence counts can be neglected for eight-fold entanglement. Therefore two-photon event count rate is $\sqrt[4]{2.8\times10^{-5}}Hz$. Detecting ten-photon entanglement requires totally five independent pairs of entangled photons to present at the same time, so the ten-photon coincidence event scales as  $(\sqrt[4]{2.8\times10^{-5}\times3600})^5=0.0568$ per hour. For ten-photon entanglement, 11 measurement settings are required according to the entanglement witness of SC state. If only 10 copies of ten-photon SC state are prepared and measured in one setting, then 110 copies of SC state are required. Therefore, the corresponding time is $(110/0.0568) hours=1.9366\times10^3 hours=80.6917 days\approx3 months$.

\section*{Entanglement witness}

 To calculate $F_{exp}$, each term in the decomposition of  $|SC\rangle\langle SC|$ has to be measured to determine its expectation value. For an eight-qubit SC state, $n = 8$, the expectation values
of all the terms on the right hand of Eq.(\ref{DEsc}) should be calculated. Specifically the total accumulated coincident counts on the i-th base is defined as $n_i$s such as $n_1$ copies of $\rho_{8photons}$ with all qubits are projected into horizontal polarization $|H\rangle$. $n_{256}$ copies of $\rho_{8photons}$ with all qubits are projected into vertical polarization $|V\rangle$. Relative frequencies on $(|H\rangle\langle H|)^{\otimes 8}$ or
$(|V\rangle\langle V|)^{\otimes 8}$ can be calculated by $n_1/(\sum_{i=1}^{2^8}n_i)$ or $n_{256}/(\sum_{i=1}^{2^8}n_i)$. To get the expectation value of the third term of Eq.(\ref{DEsc}), we have
\begin{eqnarray}
&&\frac{1}{2}[{\rm Tr}\rho_{exp}\frac{1}{8}\sum_{k=1}^{n}(-1)^{k}M^{\otimes 8}_{k\pi/8}]
\nonumber\\
&&=\frac{1}{16}[-{\rm Tr}\rho_{exp}M^{\otimes 8}_{\pi/8}+{\rm Tr}\rho_{exp}M^{\otimes 8}_{2\pi/8}-{\rm Tr}\rho_{exp}M^{\otimes 8}_{3\pi/8}\nonumber\\
&&\hspace{1.3cm}+{\rm Tr}\rho_{exp}M^{\otimes 8}_{4\pi/8}-{\rm Tr}\rho_{exp}M^{\otimes 8}_{5\pi/8}+{\rm Tr}\rho_{exp}M^{\otimes 8}_{6\pi/8}\nonumber\\
&&\hspace{1.3cm}-{\rm Tr}\rho_{exp}M^{\otimes 8}_{7\pi/8}+{\rm Tr}\rho_{exp}M^{\otimes 8}_{\pi}]\nonumber\\
&&=\frac{1}{16}\sum_{k=1}^{8}(-1)^{k}\langle M^{\otimes 8}_{k\pi/8}\rangle,
\end{eqnarray}
in which $\langle M^{\otimes 8}_{k\pi/8}\rangle$ represents the expectation of the operator $M^{\otimes 8}_{k\pi/8}$.
The estimation of expectation value of the operator
$M^{\otimes 8}_{k\pi/8}=(|+,\theta\rangle\langle +,\theta|-|-,\theta\rangle\langle -,\theta|)^{\otimes 8}$
is equivalent all the expectations of various  combinations of $|+,\theta\rangle\langle +,\theta|$ and $|-,\theta\rangle\langle -,\theta|$, due to
\begin{eqnarray}
M^{\otimes 8}_{k\pi/8}
&=&(|+,\theta\rangle\langle +,\theta|-|-,\theta\rangle\langle -,\theta|)^{\otimes 8}\nonumber\\
&=&(|+,\theta\rangle\langle +,\theta|)^{\otimes 8}-(|+,\theta\rangle\langle +,\theta|)^{\otimes 7}(|-,\theta\rangle\langle -,\theta|)\nonumber\\
&&+ .... +(|-,\theta\rangle\langle -,\theta|)^{\otimes 8}.\label{A2}
\end{eqnarray}
There are 256 terms in all for a fixed $\theta$.
 When the number of copies of state that projected into different combinations of bases $|+,\theta\rangle$ and $|-,\theta\rangle$ are collected, the copy numbers corresponding to $M^{\otimes 8}_{k\pi/8}$ are calculated from Eq.(\ref{A2}). Thus, the $\langle M^{\otimes 8}_{k\pi/8}\rangle$ can be evaluated \cite{43,5}. From
these measurements, the expectations of different terms appearing in the decomposition of the SC state entanglement witness are obtained.

\section*{Standard deviation of fidelity}

The error is calculated from Poisson distribution in Fig.2 and Fig.3 in Ref.\cite{5} for each term of Eq.(\ref{DEsc}).

Based on the experimental data of Ref.\cite{2}, the all eightfold coincidences are mainly projected into $(|H\rangle\langle H|)^{\otimes8}$ or $(|V\rangle\langle V|)^{\otimes8}$ in $S_{8photons,1}$ setting. When the state is projected into the setting of horizontal or vertical polarization, the copy of $\rho_{8qubits}$ projected into $(|H\rangle\langle H|)^{\otimes8}$ is 148, the number of copies of $\rho_{8qubits}$ is projected into $(|V\rangle\langle V|)^{\otimes8}$ is 136.
The summation of total number is 68 for the case that eight qubits are projected into other bases in $S_{8photons,1}$. Therefore, the ratio $P_1$ between the number that projected into $(|H\rangle\langle H|)^{\otimes8}$ or $(|V\rangle\langle V|)^{\otimes8}$ and the total number is (148+136)/(136+148+68)=(148+136)/352=284/352=0.8068, the ratio ($1-P_1$) between the copy that some qubits are projected into the horizontal polarization $|H\rangle$ and some are projected into vertical polarization $|V\rangle$ and the total copy of $\rho_{8photons}$ for this setting is 1-0.8068=0.1932. The smaller value in $P_1$ and $1-P_1$ is defined as $\widetilde{P_1}$.      Similarly, according to Eq.(\ref{A2}). A smaller value between $\langle M_{(j-2)\pi/8}^{\otimes8}\rangle$ and $1-\langle M_{(j-2)\pi/8}^{\otimes8}\rangle$ is chosen as $\widetilde{P_j}$, $j=2,3,\cdots,9$. Therefore, $\widetilde{P_2}=40/200=0.2$, $\widetilde{P_3}=20/107= 0.1869$, $\widetilde{P_4}=20/100=0.2$, $\widetilde{P_5}=21/110= 0.1909$, $\widetilde{P_6}=23/111= 0.2072$, $\widetilde{P_7}=19/106= 0.1792$, $\widetilde{P_8}=24/116= 0.2069$, $\widetilde{P_9}=20/103= 0.1942$, in which the largest ratio is 0.2072.

 When the number of copy of state in the whole measurement time are large and the relative frequency that the copy of state projected into a base or several bases in a setting is close to 0, Poisson distribution can be approximated by binomial distribution (Page 291 of Ref.\cite{ProbabStatisc}). Notice that the Poisson distribution here is not for the entangled photons created in BBO in time scale but the distribution of number of copies of state on different measurement basis satisfied. The binomial distribution is a special case of the Poisson binomial distribution, which is a sum of $t_j$ independent non-identical Bernoulli trials \cite{9PoissonBino}. In our optimization model, binomial distribution is applied since the number of copy of state is quite large. Let $P_1$ represent the probability that the copy of state is projected into $|H\rangle\langle H|^{\otimes n}$ or $|V\rangle\langle V|^{\otimes n}$ in the setting of $S_{1}$, where $S_{1}=\{|H\rangle\langle H|^{\otimes n}, |H\rangle\langle H|^{\otimes n-1}|V\rangle\langle V|,\cdots,|V\rangle\langle V|^{\otimes n} \}$. And let $\overline{P_1}$ denote the ratio that the state collapses to other bases in $S_{1}$, hence $\overline{P_1}=1-P_1$. According to Fig.2 and Fig.3 of Ref.\cite{5}, a value in $P_1$ or $\overline{P_1}$ is close to $1$ and the other  is close to $0$ when $t_1$ is much larger than 20. It satisfies the condition that Poisson binomial distribution can be approximately replaced by binomial distribution. Since the variance of the binomial distribution is $t_1\overline{P_1}(1-\overline{P_1})$ (Page 277 of Ref.\cite{ProbabStatisc}), then variance of number of events that SC state collapses to $|H\rangle^{\otimes n}$ or $|V\rangle^{\otimes n}$ is also the same value since $t_1\overline{P_1}(1-\overline{P_1})=t_1P_1(1-P_1)$. Therefore the standard deviation is $\sqrt{t_1P_1(1-P_1)}$. Besides, the $P_1$ is defined as the ratio between the number of copies of state detected on a $(|H\rangle\langle H|)^{\otimes n}$ or $(|V\rangle\langle V|)^{\otimes n}$ basis and the total copies of state in $S_1$. Therefore the standard deviation for the relative frequency is $\sqrt{t_1P_1(1-P_1)}/t_1$, which is equal to $\sqrt{P_1(1-P_1)/t_1}$.

 From Eq.(\ref{4}) and Eq.(\ref{DEsc}),

\begin{eqnarray}
&&F_{exp}(|SC\rangle)\nonumber\\
&&= {\rm Tr}\left(\rho_{exp}|SC\rangle\langle SC|\right)\nonumber\\
&&= {\rm Tr}\left\{\rho_{exp}\left[\frac{1}{2}\left((|H\rangle\langle H|)^{\otimes n}+(|V\rangle\langle V|)^{\otimes n}+\frac{1}{n}\sum_{k=1}^{n} (-1)^{k}M^{\otimes n}_{k\pi/n}\right)\right]\right\}\nonumber\\
&&= {\rm Tr}\left\{\rho_{exp}\left[\frac{1}{2}\left((|H\rangle\langle H|)^{\otimes n}+(|V\rangle\langle V|)^{\otimes n}\right)\right]\right\}+{\rm Tr}\left\{\rho_{exp}\left[\frac{1}{2n}\sum_{k=1}^{n} (-1)^{k}M^{\otimes n}_{k\pi/n}\right]\right\}\nonumber\\
&&= {\rm Tr}\left\{\rho_{exp}\left[\frac{1}{2}\left(|H\rangle\langle H|\right)^{\otimes n}+\left(|V\rangle\langle V|\right)^{\otimes n}\right]\right\}+\sum_{k=1}^{n}{\rm Tr}\left\{\rho_{exp}\left[\frac{1}{2n}(-1)^{k}M^{\otimes n}_{k\pi/n}\right]\right\}\nonumber\\
&&= \frac{1}{2}{\rm Tr}\left\{\rho_{exp}\left[\left(|H\rangle\langle H|\right)^{\otimes n}+(|V\rangle\langle V|)^{\otimes n}\right]\right\}+\sum_{k=1}^{n}\frac{1}{2n}(-1)^{k}{\rm Tr}\left\{\rho_{exp}\left[M^{\otimes n}_{k\pi/n}\right]\right\}\nonumber\\
&&= \frac{1}{2}{\rm Tr}\left\{\rho_{exp}\left[\left(|H\rangle\langle H|\right)^{\otimes n}+(|V\rangle\langle V|)^{\otimes n}\right]\right\}\nonumber\\
&&+\sum_{k=1}^{n}\frac{1}{2n}(-1)^{k}{\rm Tr}\left\{\rho_{exp}\left[(|+,\theta\rangle\langle +,\theta|)^{\otimes n}-(|+,\theta\rangle\langle +,\theta|)^{\otimes n-1}(|-,\theta\rangle\langle -,\theta|)
+\cdots +(|-,\theta\rangle\langle -,\theta|)^{\otimes n}\right]\right\}\nonumber\\
&&= \frac{1}{2}{\rm Tr}\left\{\rho_{exp}\left[\left(|H\rangle\langle H|\right)^{\otimes n}+(|V\rangle\langle V|)^{\otimes n}\right]\right\}\nonumber\\
&&+\sum_{k=1}^{n}\frac{(-1)^{k}}{2n}{\rm Tr}\left\{\rho_{exp}\left[(|+,\theta\rangle\langle +,\theta|)^{\otimes n}+(|+,\theta\rangle\langle +,\theta|)^{\otimes n-2} \otimes(|-,\theta\rangle\langle -,\theta|)^{\otimes 2}
+\cdots+(|-,\theta\rangle\langle -,\theta|)^{\otimes n}\right]\right\}\nonumber\\
&&-\sum_{k=1}^{n}\frac{(-1)^{k}}{2n}{\rm Tr}\{\rho_{exp}[(|+,\theta\rangle\langle +,\theta|)^{\otimes n-1}(|-,\theta\rangle\langle -,\theta|)+(|+,\theta\rangle\langle +,\theta|)^{\otimes n-2}(|-,\theta\rangle\langle -,\theta|)\otimes(|+,\theta\rangle\langle +,\theta|)
+\cdots+ (|-,\theta\rangle\langle -,\theta|)^{\otimes n-1}\otimes(|+,\theta\rangle\langle +,\theta|)]\}.\label{FEXPDECOMPp}
\end{eqnarray}

Define
\begin{eqnarray}
&&P_1={\rm Tr}\left\{\rho_{exp}\left[\left(|H\rangle\langle H|\right)^{\otimes n}+(|V\rangle\langle V|)^{\otimes n}\right]\right\},\nonumber\\
&&P_j={\rm Tr}\left\{\rho_{exp}\left[(|+,\theta\rangle\langle +,\theta|)^{\otimes n}+(|+,\theta\rangle\langle +,\theta|)^{\otimes n-2} \otimes(|-,\theta\rangle\langle -,\theta|)^{\otimes 2}+\cdots+(|-,\theta\rangle\langle -,\theta|)^{\otimes n}\right]\right\},\nonumber\\
&&1-P_j={\rm Tr}\{\rho_{exp}[(|+,\theta\rangle\langle +,\theta|)^{\otimes n-1}(|-,\theta\rangle\langle -,\theta|)+(|+,\theta\rangle\langle +,\theta|)^{\otimes n-2}(|-,\theta\rangle\langle -,\theta|)\otimes(|+,\theta\rangle\langle +,\theta|)\nonumber\\
&&+\cdots+ (|-,\theta\rangle\langle -,\theta|)^{\otimes n-1}\otimes(|+,\theta\rangle\langle +,\theta|)]\}\}.\label{p1pj1pj}
\end{eqnarray}

Then Eq.(\ref{FEXPDECOMPp}) can be rewritten as

\begin{eqnarray}
&& \frac{1}{2}P_1+\sum_{k=1}^{n}\frac{(-1)^k}{2n}P_{k+1}-\sum_{k=1}^{n}\frac{(-1)^k}{2n}(1-P_{k+1})\nonumber\\
&&=\frac{1}{2}P_1+2\sum_{k=1}^{n}\frac{(-1)^k}{2n}P_{k+1}-\sum_{k=1}^{n}\frac{(-1)^k}{2n}\nonumber\\
&&=\frac{1}{2}P_1+\sum_{j=2}^{n+1}\frac{(-1)^{j-1}}{n}P_{j}-\sum_{j=2}^{n+1}\frac{(-1)^{j-1}}{2n}.
\label{FEXPDECOMP}
\end{eqnarray}

Here Eq.(\ref{A2}) is applied when $n=8$ and $k+1$ is denoted as $j$ in last second step.

According to the previous analysis, the standard deviation of $P_j$ is
\begin{eqnarray}
\sqrt{\frac{P_{j}(1-P_{j})}{t_j}},\quad j=1, 2, \cdots, 9.\label{fHVP}
\end{eqnarray}

Further considering the formula of combined standard uncertainty \cite{erroruncertainty}, the standard deviation of fidelity can be derived. We use $\Delta F_{exp}$ to represent it. Therefore,
\begin{eqnarray}
&&\Delta F_{exp}\nonumber\\
&&=\sqrt{\left[ \frac{\partial F_{exp}}{\partial P_{1}}\sqrt{\frac{P_1(1-P_1)}{t_1}} \right]^2+\sum_{j=2}^{n+1}\left[\frac{\partial F_{exp}}{\partial P_{j}}\sqrt{\frac{P_j(1-P_j)}{t_j}} \right]^2}\nonumber\\
&&=\sqrt{\frac{1}{4}\frac{P_1(1-P_1)}{t_1}+\frac{1}{n^2}\sum_{j=2}^{n+1}\frac{P_j(1-P_j)}{t_j}}.\label{deltaF}
\end{eqnarray}

\section*{Theoretical derivation of Minimum copies of multi-photon Schr\"{o}dinger's Cat state}

Let
\begin{eqnarray}
&& k_1=\frac{1}{4}P_1(1-P_1),\nonumber\\
&& \ k_j=\frac{1}{n^2}P_j(1-P_j), j=2, 3, \cdots, n+1.\nonumber \\
&& \epsilon=\epsilon_{0}^2, \label{k1j}
\end{eqnarray}

then the optimization problem is equivalent to

\begin{eqnarray}
&& {\rm Minimize_{t_1,t_2,\cdots,t_{n+1}}}\ \sum_{j=1}^{n+1}t_j\  \nonumber\\
&& {\rm subject\ to}\ \sum_{j=1}^{n+1}\frac{k_j}{t_j}\leq \epsilon, t_j>0,k_j\geq0, j=1,2,\cdots,n+1\label{optimizationp}
\end{eqnarray}

where $\epsilon$ and $k_j$ are positive real constants, $n$ is a positive integer, $t_j$ is a variable and also a positive integer. In order to solve Eq.(\ref{optimizationp}) easily, all the variables, including the number of copies of state, $t_j$, are considered as a real. The optimized number of copies of state is then rounded off to the smallest integer greater than the final real $t_j$.

 The solution of the optimization problem is assumed to satisfy $\sum_{j=1}^{n+1}\frac{k_j}{t_j}= \epsilon$. If the optimal solution is not on the boundary, it means $\sum_{j=1}^{n+1}\frac{k_j}{t_j} < \epsilon$. Appropriate reduction in the number of $t_j$ can be made, while the inequality ($\sum_{j=1}^{n+1}\frac{k_j}{t_j}\leq \epsilon$) is still satisfied. This is contradictory with the target function ``${\rm Minimize} \sum_{j=1}^{n+1}t_j$", therefore the optimal solution must exist on the bound.

The detailed process of how to find the analytical solution of Eq.(\ref{optimizationp}) will be shown below.

Now let
\begin{eqnarray}
&& t_j=\frac{1}{x_j}. \label{tjtran}
\end{eqnarray}

By substituting Eq.(\ref{tjtran}) into  Eq.(\ref{optimizationp}), we obtain
\begin{eqnarray}
&&{\rm Minimize}\ \sum_{j=1}^{n+1}\frac{1}{x_j}\  \nonumber\\
&&{\rm subject\ to} \nonumber\\
&& \sum k_jx_j=\epsilon. \label{Minxj}
\end{eqnarray}

The Lagrange multiplier method is applied to solve the problem. Since the target is the minimization of $\sum_{j=1}^{n+1}\frac{1}{x_j}$, Eq.(\ref{Minxj}) leads to
\begin{eqnarray}
L=\sum \frac{1}{x_j}-\lambda(\sum k_jx_j-\epsilon).
\end{eqnarray}
To find its minimum, partial derivative for each $x_j$ is expressed as
\begin{eqnarray}
\frac{\partial L}{\partial x_j}=-\frac{1}{x_j^{2}}-k_j\lambda=0. \label{lagr0}
\end{eqnarray}

From Eq.(\ref{lagr0}), the following equation is obtained,
\begin{eqnarray}
x_j=\sqrt{\frac{-1}{k_j\lambda}}. \label{lagr1}
\end{eqnarray}

The constraint of Eq.(\ref{Minxj}) is
\begin{eqnarray}
\sum k_jx_j=\epsilon. \label{lagr2}
\end{eqnarray}

From Eq.(\ref{lagr1}) and Eq.(\ref{lagr2}), $\lambda$ is given by
\begin{eqnarray}
\lambda=-\left(\frac{\sum\sqrt{k_j}}{\epsilon}\right)^2. \label{lagr3}
\end{eqnarray}

By substituting Eq.(\ref{lagr3}) into Eq.(\ref{lagr1}), we arrive at

\begin{eqnarray}
x_j=\frac{\epsilon}{\sqrt{k_j}(\sum_{j=1}^{n+1}\sqrt{k_j})}.\label{xjee}
\end{eqnarray}

Rewriting Eq.(\ref{xjee}) using Eq.(\ref{tjtran}), we have

\begin{eqnarray}
t_j=\frac{\sqrt{k_j}(\sum_{j=1}^{n+1}\sqrt{k_j})}{\epsilon}.\label{tjjjj}
\end{eqnarray}

By substituting Eq.(\ref{k1j}) into Eq.(\ref{tjjjj}), Eq.(\ref{optimalt}) is obtained. The optimal results of Eq.(\ref{optimalt}) can be compared with experiment when the same coefficient $P_1$ and $P_j$s are substituted.

\section*{Theoretical derivation of minimum number of copies of a state in quantum-state tomography}

 According to Born's rule,
\begin{eqnarray}
&&P_{\mu,\nu}={\rm Tr}(M_{\mu,\nu}\rho_0),\nonumber\\
&&f_{\mu,\nu}={\rm Tr}(M_{\mu,\nu}\rho),\nonumber\\
&&\mu=1,2,\cdots,d,\nu=1,2,\cdots,n_s,\label{pifi}
\end{eqnarray}
where $\mu$ distinguishes different measurement operators in the same setting, $d$ is the dimension of density matrix and $P_{\mu,\nu}$ is the probability when $\rho_0$ is measured by operator $M_{\mu,\nu}$. Namely, $\rho\in C^{d\times d}$, then
\begin{eqnarray}
&&\rho_0=\sum_{\mu,\nu}P_{\mu,\nu}M_{\mu,\nu},\nonumber\\
&&\rho=\sum_{\mu,\nu}f_{\mu,\nu}M_{\mu,\nu}.
\end{eqnarray}
Therefore one has
\begin{eqnarray}
&&\rho-\rho_0=\sum_{\mu,\nu}(f_{\mu,\nu}-P_{\mu,\nu})M_{\mu,\nu}.
\end{eqnarray}

Consider $M_{\mu,\nu}$ is orthogonal to each other, then
\begin{eqnarray}
&&||\rho-\rho_0||_F\nonumber\\
&&=\left[{\rm Tr}\left((\rho-\rho_0)(\rho-\rho_0)^*\right)\right]^{1/2}\nonumber\\
&&=\left[\sum_{\mu,\nu}(f_{\mu,\nu}-P_{\mu,\nu})^{2}{\rm Tr}(M_{\mu,\nu}M_{\mu,\nu}^*)\right]^{1/2}\nonumber\\
&&=\left[\sum_{\mu,\nu}\left(\sqrt{\frac{f_{\mu,\nu}(1-f_{\mu,\nu})}{T_{\mu,\nu}}}\right)^{2}{\rm Tr}(M_{\mu,\nu}M_{\mu,\nu}^*)\right]^{1/2}.\label{rhoF}
\end{eqnarray}

 In the last step of Eq.(\ref{rhoF}), standard deviation of binomial distribution is applied, see the first paragraph of ``Standard deviation of fidelity" in appendix for detail. When the measurement operator $M_{\mu,\nu}$ belongs to the same setting $\nu$, they have the identical number of copies $T_{\mu,\nu}$ of $\rho_0$, i.e.$T_{1,\nu}=T_{2,\nu}=\cdots=T_{d,\nu}$. We denote them to be $T_\nu$.
The model of Eq. (\ref{Optimizationtomography}) is equivalent to

\begin{eqnarray}
&& {\rm Minimize}\ \sum_{\nu=1}^{n_s}T_{\nu}\  \nonumber\\
&& {\rm subject\ to}\ \nonumber\\
&&\left[\sum_{\nu=1}^{n_s}\sum_{\mu=1}^{d}\left(\sqrt{\frac{f_{\mu,\nu}(1-f_{\mu,\nu})}{T_{\nu}}}\right)^{2}{\rm Tr}(M_{\mu,\nu}M_{\mu,\nu}^*)\right]^{1/2} \nonumber\\
&&\leq \epsilon_0.\label{Optimizationtomography2}
\end{eqnarray}

It is easy to find that the target of Eq.(\ref{Optimizationtomography2}) is similar to Eq.(\ref{Optimizationproblem}) except the larger required number of settings and different coefficients.

Let $\sum_{\mu=1}^{d}[f_{\mu,\nu}(1-f_{\mu,\nu}){\rm Tr}(M_{\mu,\nu}M_{\mu,\nu}^*)]$ be $k_\nu$ and $\epsilon=\epsilon_0^2$, then the model has the following form
\begin{eqnarray}
&& {\rm Minimize}\ \sum_{\nu=1}^{n_s}T_{\nu}\  \nonumber\\
&& {\rm subject\ to}\ \sum_{\nu=1}^{n_s}\frac{k_\nu}{T_\nu}\leq \epsilon, T_\nu>0,k_\nu\geq0, \nu=1,2,\cdots,n_s.\label{optimizationp2}
\end{eqnarray}

Obviously, it has a similar form to Eq.(\ref{optimizationp}); therefore the solution is the same as that of Eq.(\ref{optimalt})
\begin{eqnarray}
T_\nu=\frac{\sqrt{k_\nu}(\sum_{\nu=1}^{n_s}\sqrt{k_j})}{\epsilon}.\label{optimalt2}
\end{eqnarray}

If $M_{\mu,\nu}$ is non orthogonal with each other, then
\begin{eqnarray}
&&||\rho-\rho_0||_F\nonumber\\
&&=\left\{{\rm Tr}\left[(\rho-\rho_0)(\rho-\rho_0)^*\right]\right\}^{1/2}\nonumber\\
&&=\left[\sum_{\mu,\nu,\mu^{'},\nu^{'}}(f_{\mu,\nu}-P_{\mu,\nu})(f_{\mu^{'},\nu^{'}}-P_{\mu^{'},\nu^{'}}){\rm Tr}(M_{\mu,\nu}M_{\mu^{'},\nu^{'}}^*)\right]^{\frac{1}{2}}\nonumber\\
&&=\left[\sum_{\mu,\nu,\mu^{'},\nu^{'}}\sqrt{\frac{f_{\mu,\nu}(1-f_{\mu,\nu})}{T_{\mu,\nu}}}\sqrt{\frac{f_{\mu^{'},\nu^{'}}(1-f_{\mu^{'},\nu^{'}})}{T_{\mu^{'},\nu^{'}}}} {\rm Tr}(M_{\mu,\nu}M_{\mu^{'},\nu^{'}}^*)\right]^{1/2}.
\end{eqnarray}
Applying the similar substitutions as used in the orthogonal case, we have
 \begin{eqnarray}
&&{\rm Minimize}\ \sum_{\nu=1}^{n_s}T_{\nu}\  \nonumber\\
&&{\rm subject\ to}\ \sum_{\nu,\nu^{'}=1}^{n_s}\frac{k_{\nu,\nu^{'}}}{\sqrt{T_{\nu}}\sqrt{T_{\nu^{'}}}}\leq \epsilon, \nonumber\\ &&T_{\nu}>0,T_{\nu^{'}}>0,k_{\nu,\nu^{'}}\geq0, \nonumber\\
&&\nu=1,2,\cdots,n_s, \nu^{'}=1,2,\cdots,n_s.\label{Optimizationtomography3}
\end{eqnarray}
Substitute $q_{\nu}=1/T_{\nu}$, the constraint in the optimization becomes
\begin{eqnarray}
&&\sum_{\nu,\nu^{'}=1}^{n_s}\frac{k_{\nu,\nu^{'}}}{\sqrt{T_{\nu}}\sqrt{T_{\nu^{'}}}}\nonumber\\
&&=\sum_{\nu,\nu^{'}=1}^{n_s}k_{\nu,\nu^{'}}\sqrt{q_{\nu}}\sqrt{q_{\nu^{'}}}\nonumber\\
&&\leq \sum_{\nu,\nu^{'}=1}^{n_s}k_{\nu,\nu^{'}}(q_{\nu}+q_{\nu^{'}})/2\nonumber\\
&&\leq \sum_{p=1}^{n_s}\left[\sum_{\nu^{'}=1}^{n_s}k_{p,\nu^{'}}+\sum_{\nu=1}^{n_s}k_{\nu,p}\right]q_{p}/2\nonumber\\
&&\leq \epsilon.
\end{eqnarray}
Therefore, the non orthogonal case has a similar result with the orthogonal one Eq.(\ref{optimizationp2}) except the coefficient is different.

\section*{Acknowledgements}

The authors would like to thank Prof. Jian-wei Pan, Prof. Chaoyang Lu and other members of their group for providing their experimental data. The authors
also would like to thank Dr.Wenkai Yu and Dr.Yulong Liu for helpful discussions. We further express our sincere gratitude to Prof.Adam Liwo and Mr. Waqas Mahmood for making revisions of article. The authors would like to greatly thank Prof. Mo-Lin Ge for the valuable discussions. This work is supported by NSF of China with the Grant No. 11275024 and No.11475088. Additional support was provided by the Ministry of Science and Technology of China (2013YQ030595-3, and 2013AA122901).

\section*{Author contributions statement}

Y.L. constructed the model and performed the numerical simulations, Q.Z. supervised the research. All
authors contributed to the preparation of this manuscript.

\section*{Additional information}
\textbf{Competing financial interests}: The authors declare no competing financial interests.

\newpage

\end{document}